\documentclass[10pt]{iopart}
\usepackage{iopams}  
\usepackage[english]{babel}
\expandafter\let\csname equation*\endcsname\relax
\expandafter\let\csname fl\endcsname\relax
\usepackage{amsfonts,amssymb,amstext,amsthm,amscd,amsbsy}
\expandafter\let\csname endequation*\endcsname\relax
\expandafter\let\csname endfl\endcsname\relax
\usepackage{mathtools}
\usepackage[a4paper,top=2.5cm,bottom=2.5cm,left=2.5cm,right=2.5cm]{geometry}
\usepackage{breakcites}
\usepackage{multirow}
\usepackage{booktabs}
\usepackage{mathrsfs}
\usepackage{dsfont}

\usepackage{graphicx}
\usepackage{wrapfig}
\usepackage{indentfirst}
\usepackage{latexsym}
\usepackage{sidecap}
\usepackage{booktabs}
\usepackage{hyperref}

\usepackage{lipsum}
\usepackage{cleveref}
\usepackage[dvipsnames]{xcolor}

\newcommand\myshade{85}
\colorlet{mylinkcolor}{violet}
\colorlet{mycitecolor}{YellowOrange}
\colorlet{myurlcolor}{RoyalBlue}

\hypersetup{
    colorlinks=true,
    linkcolor=myurlcolor,
    citecolor=mycitecolor!\myshade!black,
    filecolor=magenta,      
    urlcolor=magenta,
}

\usepackage{mathrsfs}
\usepackage{textcomp}
\usepackage{braket}
\usepackage{colortbl}
\usepackage{bbm}
\usepackage{comment}
\usepackage{textcomp}
\usepackage{braket}
\usepackage{colortbl}
\usepackage{bbm}

\usepackage{enumitem}
\usepackage{musicography}
\usepackage{caption}
\usepackage{subcaption}

\usepackage{tikz}
\usepackage{pgfplots}
\usetikzlibrary{matrix}
\usepackage{comment}

\usepackage{graphicx}
\usepackage{tikz}
\usetikzlibrary{positioning, fit, arrows.meta, shapes}
\def\layersep{2.5cm}

\newcommand{\RBE}[1]{RBE$_{#1}\,$}
\newcommand{\AB}{$\alpha_\gamma/\beta_\gamma\,$}
\newcommand{\BA}{$\beta_\gamma/\alpha_\gamma\,$}
\newcommand{\kev}{keV$/\mu$m $\,$}
\newcommand{\AG}{$\alpha_\gamma\,$}
\newcommand{\BG}{$\beta_\gamma \,$}
\newcommand{\mev}{MeV$/$u $\,$}

\newcommand{\red}[1]{\textcolor{black}{#1}}

\usepackage{fancyhdr}
\usepackage{lipsum}% just to generate text for the example

\begin{document}

\title[An AI-based model for cell killing prediction]{An Artificial Intelligence-based model for cell killing prediction: development, validation and explainability analysis of the ANAKIN model}

\author{Francesco G. Cordoni$^{1,2}$, Marta Missiaggia$^{2,3}$, Emanuele Scifoni$^{2}$ and Chiara La Tessa$^{2,3,4}$}

\renewcommand{\thefootnote}{\fnsymbol{footnote}}
\footnotetext{{\scriptsize $^{1}$ Department of Civil, Environmental and Mechanical Engineering, via Mesiano 77, 38123, Trento, Italy}}
\footnotetext{{\scriptsize $^{2}$ Trento Institute for Fundamental Physics and Application (TIFPA), via Sommarive 15, 38123, Trento, Italy}}
\footnotetext{{\scriptsize $^{3}$ Department of Radiation Oncology, University of Miami Miller School of Medicine, 33136, Miami FL, USA}}
\footnotetext{{\scriptsize $^{4}$ Department of Physics, via Sommarive 14, 38123, Trento, Italy}}
\footnotetext{{\scriptsize E-mail addresses: francesco.cordoni@unitn.it, mxm194331@miami.edu, emanuele.scifoni@tifpa.infn.it, cxl1666@miami.edu}}
\vspace{10pt}

\begin{abstract}
The present work develops ANAKIN: an \textit{Artificial iNtelligence bAsed model for \red{(radiation induced)} cell KIlliNg prediction}. ANAKIN is trained and tested over 513 cell survival experiments \red{with different types of radiation} contained in the publicly available PIDE \red{database}. We show how ANAKIN accurately predicts several relevant biological endpoints over a wide broad range on ions beams and for a high number of cell--lines. We compare the prediction of ANAKIN to the only two radiobiological model for RBE prediction used in clinics, that is the \textit{Microdosimetric Kinetic Model} (MKM) and the \textit{Local Effect Model} (LEM \red{version III}), showing how ANAKIN has higher accuracy over the all \red{considered} biological endpoints. At last, via modern techniques of \textit{Explainable Artificial Intelligence} (XAI), we show how ANAKIN predictions can be \red{understood} and explained, highlighting how ANAKIN is in fact able to reproduce relevant well-known biological patterns, such as the overkilling effect.
\end{abstract}

\cleardoublepage

\tableofcontents

\cleardoublepage

\section{Introduction}\label{SEC:Intro}

In the last decades, radiotherapy (RT) has \red{increasingly} proven to be an extremely effective cure against cancer. Within RT, particle therapy (PT), has been emerging ~\cite{durante2019charged}, and at the end of 2021, about 325.000 patients have been treated worldwide with PT, of which close to 280.000 with protons and about 42.000 with carbon ions \cite{PTCOG}. Furthermore, other ions have been recently gaining attention~\cite{Rovituso2017}: in 2021 the first patient was (re-)treated with helium \cite{mairani2022roadmap} at the Heidelberg Ion Therapy center (HIT) in Germany, while \red{perspective studies are looking into the possible using of oxygen} \cite{kurz2012first,sokol2017oxygen}.

The physical rationale of using hadrons in cancer treatment is their characteristic energy loss mechanisms, that result in concrete biological advantages compared to photons, such an increased tumor control and a greater sparing of normal tissues, with a consequently lower risk of toxicity.

Despite the theoretical superior physical properties of hadrons compared to photons, further research is critical for increasing the PT application in the clinic. A correct and accurate estimation of radiation-induced biological damage remains one of the major limitations to the full exploitation this treatment modality. The key quantity used to describe the radiation effectiveness in inducing a specific damage is the \textit{Relative Biological Effectiveness} (RBE), which is defined as the ratio between the dose delivered by a given radiation and the dose delivered by the reference radiation yielding the same biological effect:
\[
RBE= \left . \frac{D_{\mbox{reference}}}{D_{\mbox{radiation}}} \right |_{\mbox{isoeffect}}\,.
\]

RBE allows to quantify how much more lethal a certain radiation is compared to the reference radiation, usually X-rays, and is used in \textit{Treatment Planning Systems} (TPS) to calculate the biological dose, namely the physical dose multiplied by the RBE. For this reason, over the last decades a plethora of mathematical mechanistic models, \cite{hawkins1994statistical,inaniwa2018adaptation,kase2006,bellinzona2021linking,friedrich2013systematic,friedrich2013local,elsasser2010quantification, mcmahon2021mechanistic, vassiliev2012formulation,Vas3,tobias1985repair,tobias1980repair,kellerer1974theory,kellerer1978generalized,cordoni2021cell,cordoni2021generalized,manganaro2017monte}, as well as data-driven phenomenological models, \cite{tilly2005influence,mcnamara2015phenomenological,chen2012empirical,carabe2012range,wilkens2004phenomenological,mairani2017phenomenological} have been developed to estimate RBE based on biological as well as physical quantities. At the base of most models is the linear-quadratic (LQ) behavior of the cell survival logarithm with respect to the imparted dose:
\[
S(D) = e^{-\alpha D - \beta D^2}\,,
\]
where $\alpha$ and $\beta$ are some specific parameters that depend on both biological \red{e.g. tissue type)} and physical \red{(e.g radiation quality)} variables \cite{mcmahon2018linear}.

Currently, a constant RBE of 1.1 is conservatively used in proton therapy, although evidences show its variability, especially in the distal region \cite{Paganetti2002_a,Paganetti2014,Paganetti2018,missiaggia2020microdosimetric,missiaggia2022investigation}. For carbon and helium ions, the RBE variations \red{across the irradiation field} are significant enough that a constant value cannot be used. Currently, two radiobiological models are currently used to predict RBE in clinical practice: (i) the \textit{Microdosimetric Kinetic Model} (MKM) \cite{inaniwa2010treatment,inaniwa2018adaptation,bellinzona2021linking}, and (ii) the \textit{Local Effect Model} (LEM) \cite{friedrich2013local,elsasser2010quantification,pfuhl2022comprehensive}. Both models have been vastly tested against in vitro and in vivo data \cite{pfuhl2022comprehensive,inaniwa2018adaptation}, but the outcomes have not indicated a clear superiority of one model to the other. In addition, significant differences in the prediction of RBE across models are evident so that, at present days, the use in clinical practice of a variable RBE is highly subject to the model chosen, \cite{missiaggia2022investigation,missiaggia2020microdosimetric,giovannini2016variable,bertolet2021implementation}.

The lack of a robust and generalized model for predicting RBE hinder the full exploitation of PT, including the use of ions heavier than carbon, such as oxygen, to successfully treat radio-resistant tumors, \cite{boulefour2021review}, or multi--ion therapy, which is nowadays accessible from the technical point of view~\cite{ebner2021emerging}.

Furthermore, although some RBE models have a general mathematical formulation, their implementation in the TPS, especially for inverse planning, requires a heavy calculation effort. This issue is usually overcome both by using look up tables and by making specific assumptions \cite{inaniwa2018adaptation}, such as physical or biological approximations, which clearly limit the model generality and affects its RBE prediction accuracy.

Aiming at deriving a general model able to accurately predict RBE across a wide range of physical and biological variables, we developed ANAKIN (\textit{an Artificial iNtelligence bAsed model for \red{(radiation induced)} cell KIlliNg prediction}), a new general AI-driven model for predicting cell survival and RBE. \textit{Machine Learning} (ML) and \textit{Deep Learning} (DL) algorithms have recently started to gain attention in the medical physics community with applications on imaging \cite{sahiner2019deep}, fast dose estimation \cite{gotz2020deep}, Monte Carlo simulation \cite{sarrut2021artificial}, and particle tracking \cite{missiaggia2022exploratory} have been published. However, only \cite{papakonstantinou2021using} apply ML for predicting radiation induced biological quantities, conducting a study on induction of DNA damage and its complexity, but no analysis on RBE is performed.

ANAKIN is composed by various ML and DL--based modules, each with a specific tool, and interconnected to each other. The model considers both physical variables such as the kinetic energy of the incident beam or also the \textit{Linear Energy Transfer} (LET), that is the amount of energy that a particle transfers to the material traversed per unit distance, \cite{durante2016nuclear}, and biological variables, such as the $\alpha$ and $\beta$ values for the reference radiation \red{response}. To make the model as general as possible, we trained it on cell survival data for 20 cell-lines widely used in radiobiology and 11 different ion types all available on the \textit{Particle Irradiation Data Ensemble} (PIDE)  \cite{friedrich2013systematic,friedrich2021update}. Together with particles of interest for clinical applications, we also included in the training process heavier ions, including iron. This choice extends the application of ANAKIN to other research fields, such as radiation protection in space.  
To verify ANAKIN predictions and assess its accuracy, we randomly divided the data available in PIDE into two sets, one for training and one for testing. Therefore all results reported in the present work refer to the test set, that consists entirely of experiments that have not been included into the training set. 

\textit{Artificial Intelligence} (AI) has had a disruptive impact both in the research field and in real-life applications. The potential of modern and advanced \textit{Machine Learning} (ML) and \textit{Deep Learning} (DL) algorithms have started to gain attention in the medical physics community, where several research papers on application of DL to imaging, \cite{sahiner2019deep}, fast dose estimation, \cite{gotz2020deep}, Monte Carlo simulation, \cite{sarrut2021artificial}, and particle tracking, \cite{missiaggia2022exploratory}, have appeared. Quite surprisingly, to the best of our knowledge, the only results in literature that use ML to predict radiation induced biological quantities is \cite{papakonstantinou2021using}, where the authors conduct a study on induction of DNA damage and its complexity, but no analysis on RBE is performed \cite{davidovic2021application}.

ML and DL has shown to be an extremely powerful, accurate and flexible tool to extract information and hidden relations as well as to predict the most likely outcome based on data of possibly different nature, \cite{ongsulee2018big,khalid2007machine,shwartz2022tabular}. Moreover, a\red{n excellent, systematic and}  comprehensive data collection of cell survival experiments exists and is publicly available, the \textit{Particle Irradiation Data Ensemble} (PIDE), \cite{friedrich2013systematic,friedrich2021update}.

ANAKIN is constituted by various ML and DL--based modules, each with a specific task, and interconnected to each other. Two different tree-based models, namely the \textit{Random Forest} (RF) \cite{ho1995random,ho1998random}, and the \textit{Extreme Gradient Boosting} (XGBoost) \cite{Chen2016,ChenG16} algorithms are used to predict cell survival for a wide variety of radiation and cell--lines. It is worth stressing that the final goal of ANKIN is to develop a robust and accurate model that is able to predict cell survival in the most general possible conditions. ANAKIN is trained to predict cell survival for 20 widely used cell-lines and for 11 different ions type. Concerning this last point, despite the driving motivation is HT, many different ions, such as iron which is beyond the possible application in clinic, are included into the model. This make ANAKIN extremely general so that possible future application in space radioprotection are also \red{envisaged}.

ANAKIN is trained on the PIDE. It is worth stressing that, in order to be as more realistic as possible, experiments contained in the PIDE dataset \red{are} divided into a training set and a testing set. Therefore all results reported in the present work refer to the test set, that consists entirely of experiments that have not been included into the training set. This means that ANAKIN is asked to predict the cell survival for experiments that has never been seen before. Besides the already mentioned variables, ANAKIN considers both physical variables such as the kinetic energy of the incident beam or also the \textit{Linear Energy Transfer} (LET), that is the amount of energy that a particle transfers to the material traversed per unit distance, \cite{durante2016nuclear}, and biological variables, such as the $\alpha$ and $\beta$ values for the reference radiation \red{response}. 

ANAKIN is tested over several endpoints and metrics to establish the actual accuracy of its predictions. Further, ANAKIN predictions are compared with the MKM and the LEM, which are the only two radiobiological models currently used in the clinic. Regarding the LEM results, an extremely well-done and extensive analysis of the LEM \red{has become available very recently} \cite{pfuhl2022comprehensive}. As a matter of a fact, much of the analysis conducted in the current paper has been explicitly inspired by \cite{pfuhl2022comprehensive}. In this direction, it must be stressed that, in the current paper, the version LEM III is used since the LEM IV is not currently implemented in the \textit{survival toolkit}\red{and thus, the presented comparisons could not be translated to the state of the art version of the latter code}
It is clear that the results reported in \cite{pfuhl2022comprehensive} on the LEM IV are more accurate that the one reported in the current research using the LEM III, so this fact must be take into account.

Finally, the current work further aims at demystifying the erroneous myth that ML and DL models are obscure black-box model and whose predictions cannot be interpreted. If this argument can in fact \red{be} partially correct for extremely deep and sophisticated NN that have been built mostly in the field of the \textit{Reinforcement Learning}, \red{the same} cannot be said for the vast majority of ML and DL developed in the last years. In fact, on one side, it must be said that some ML models, such as for instance tree-based models, are interpretable by nature and, on the other side, recently a huge attention has been posed to the development of mathematical techniques aiming at explaining ML and DL models that are not of easy interpretation; such area of research is known as \textit{Explainable AI} (XAI) \cite{gunning2019xai}.

The goal of the present research is to:
\begin{description}
\item[(i)] develop for the first time a general AI-driven model to predict cell survival probability over a wide range of biological cell-lines and physical irradiation conditions;\\
\item[(ii)] compare ANAKIN with the two radiobiological models used in the clinic (MKM and LEM);\\
\item[(iii)] show that ML- and DL-based models are not only accurate, but can also help in gaining new knowledge and understanding in radiobiology and medical physics.
\end{description}

\section{Material and methods}\label{SEC:MM}
\subsection{The dataset}

The development, training and verification of ANAKIN are based on data from PIDE \cite{friedrich2013systematic,friedrich2021update}.

The MKM and LEM predictions are computed via the \textit{survival toolkit}, \cite{manganaro2018survival,survival}. This toolkit is a open source implementation that has been checked to be coherent with published results of the models, but nonetheless differences with the most advanced versions of the two formalism may arise. Unfortunately, to date no extensive and qualitative estimation of the MKM predictions over many cell-lines exists so that we could only rely on the \textit{survival toolkit}.
 %For detailed explanation of variables as well as details on the experiments included into the dataset we refer to \cite{friedrich2013systematic,friedrich2021update}. Nonetheless, few modification on the dataset has been performed before ANAKIN could be trained.

The PIDE database contains a series of cell survival experiments, conducted over a multitude of different irradiation conditions and cell-lines. In addition to the original data, a set of LQ parameters are calculated for each experiment and is also reported. Following \cite{pfuhl2022comprehensive}, ANAKIN is thus trained over the exponential linear-quadratic fit on cell survival experiments. This is done as many experiments contained in the PIDE clearly shows anomalous variability in the reported survival fraction. Experiments reporting less than 3 measurement points are removed from the dataset, because at least 3 values are needed to fit an LQ curve. The dataset obtained from PIDE is then divided into two subclasses, one for training ANAKIN and one for testing its predictions. The selection is done so that each subset contains a sufficient amount of data for each cell lines and ions to be statistically significant.

Unlike \cite{pfuhl2022comprehensive}, ANAKIN is trained on both monoenergetic and SOBP ion beams, and for this reason a specific variable is added to the data to specify the irradiation condition.

After applying all the selection criteria described above, the resulting dataset contains 513 experiments, including 20 cell-lines and 11 ion types, of which 333 were randomly assigned for training and the remaining 180 for testing. Figure \ref{FIG:PIDE} gives an overall point of view on the number of considered experiments for each cell--lines as well as ion type.

At the end of the cleaning of the data we are left with 513 experiments, 333 experiments are randomly selected for training whereas the remaining 180 are used to test. It is worth stressing that the division between train and test has been performed over the experiments, meaning that ANAKIN is tested on experiments that have never been seen before by the model. All result that are shown in the present paper refer to the test set, so that ANAKIN test reflects a realistic situation in which ANAKIN should predict the cell survival of an \red{experiment} or a real situation that has never seen before. 

ANAKIN takes as input 14 variables of both physical and biological parameters, either continuous, such as LET or energy, or discrete, such as ion type or cell--lines. The full list is reported in Table \ref{TAB:PIDE}. \red{Variables} names as reported in Table \ref{TAB:PIDE} are taken from the PIDE and described in \cite{friedrich2013systematic,friedrich2021update}. The only variable that has been added to the dataset is the square of the dose, named \textit{Dose2}. The choice of considering the square of the dose is motivated by the well-know linear and quadratic form for the logarithm of the survival. It is worth noticing that also \AG and \BG values are passed as input to ANAKIN.

\begin{table}[]
\begin{center}
\begin{tabular}{|l|l|}
\hline
\multicolumn{1}{|c|}{\textbf{\begin{tabular}[c]{@{}l@{}}Physical \\ variables\end{tabular}}} & \multicolumn{1}{c|}{\textbf{\begin{tabular}[c]{@{}l@{}}Biological \\ variables\end{tabular}}} \\ \hline \hline
Dose                                                                                         & Cells                                                                                         \\
Dose2                                                                                        & CellClass                                                                                     \\
LET                                                                                          & CellOrigin                                                                                    \\
Energy                                                                                       & CellCycle                                                                                     \\
Ion                                                                                          & DNAContent                                                                                    \\
Charge                                                                                       & ax                                                                                            \\
IrradiationConditions                                                                        & bx                                                                                           \\\hline
\end{tabular}\caption{List of all variables used as input to ANAKIN. The names are described according to PIDE documentation  \cite{friedrich2013systematic,friedrich2021update} with the exception of Dose2, which represents the square of the dose value and has been introduced in this work.}\label{TAB:PIDE}
\end{center}
\end{table}

\begin{comment}
\begin{figure}
    \centering
 \includegraphics[width=.9\columnwidth]{Figure/PIDE.png}
        \caption{Number of experiments for each ion type, reported in the horizontal axis, and cell-line, reported in the vertical axis; the size refers to the number of experiments.}\label{FIG:PIDE}
\end{figure} 
\end{comment}

\subsection{Machine and Deep Learning models}

\begin{figure}
 \begin{subfigure}[b]{\textwidth}
 \includegraphics[width=.9\columnwidth]{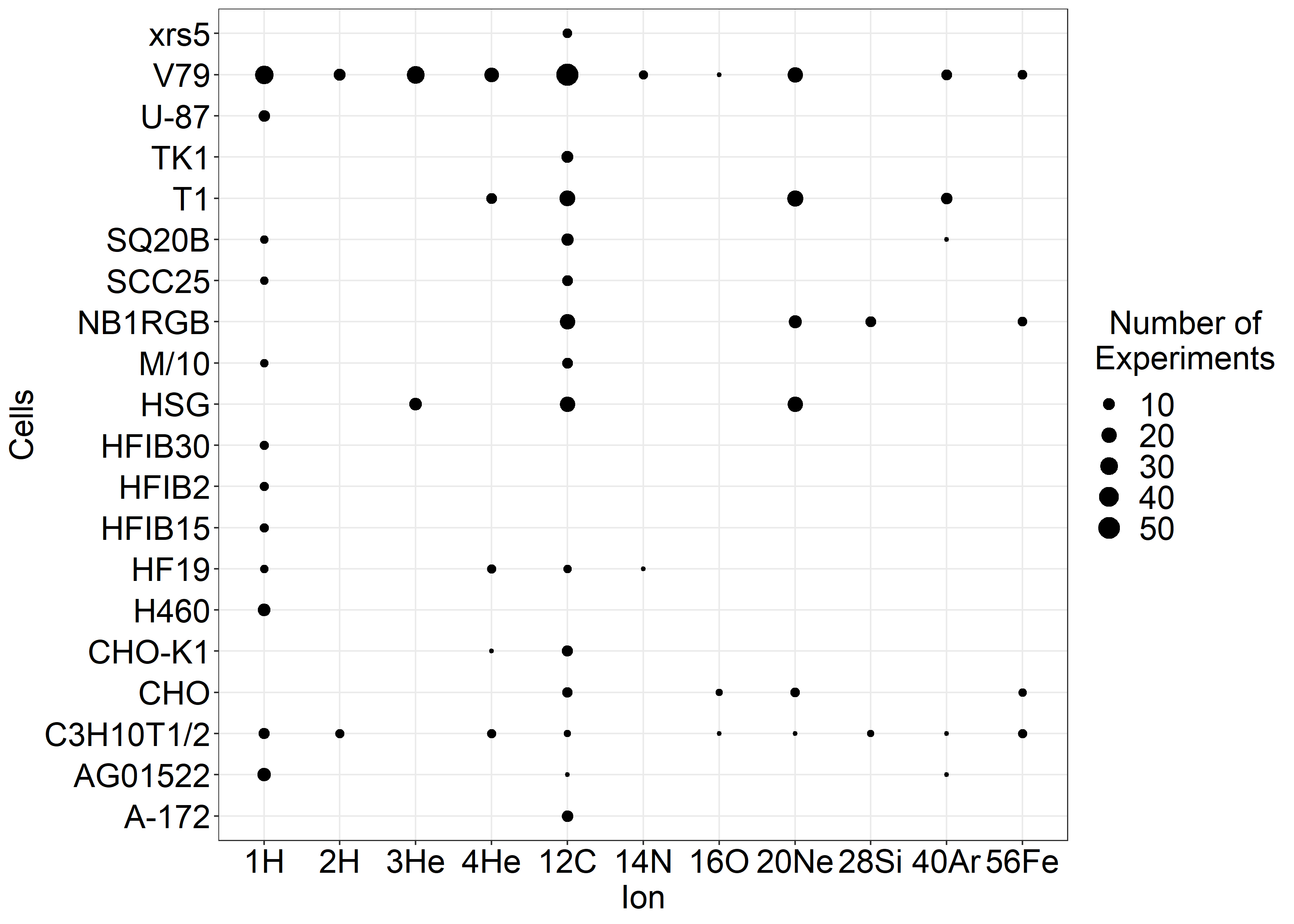}
        \caption{Number of experiments for each ion type, reported in the horizontal axis, and cell-line, reported in the vertical axis; the size refers to the number of experiments.}\label{FIG:PIDE}
\end{subfigure}
\par\bigskip
 \begin{subfigure}[b]{\textwidth}
\includegraphics[width=.9\columnwidth]{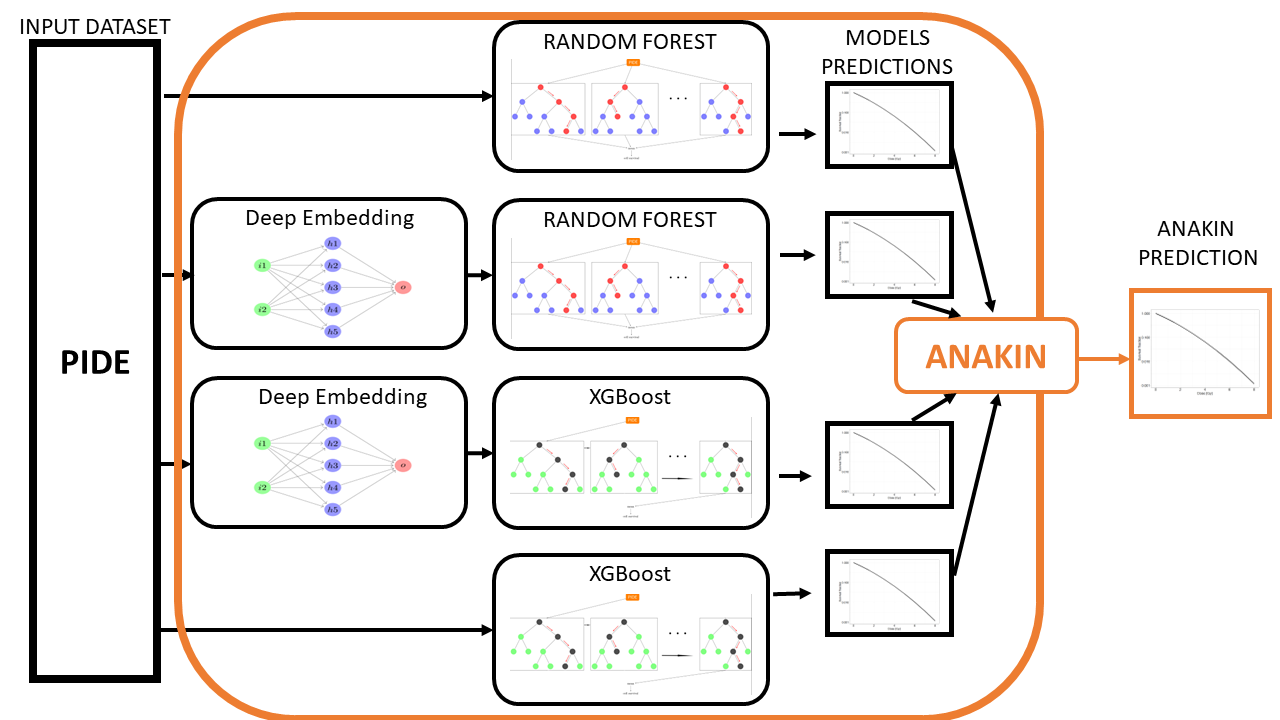}
        \caption{A schematic representation of ANAKIN. Data from PIDE are input into two types of tree-based models (RF and XGBoost), either directly or after being processed with a Deep Embedding. All four models predict cell survival, and the values are used as input for ANAKIN, that opportunely combines the predictions to provide a final cell survival output.}\label{FIG:AS}
\end{subfigure}
\caption{ANAKIN input dataset and workflow.}\label{FIG:PIDEAll}
\end{figure}

ANAKIN is an ensemble AI model composed of ML and DL modules, each with a different task, that together predict the cell-survival probability. A schematic representation of ANAKIN is shown in Figure \ref{FIG:AS}. Four different tree-based models are trained on the PIDE: two \textit{Random Forest} (RF) \cite{ho1998random,ho1995random}, and two \textit{Extreme Gradient Boosting} (XGBoost) \cite{Chen2016,ChenG16}. PIDE data are directly used as input for one RF and one XGBoost models, while for the other two they are first processed with the Deep Embedding \cite{guo2016entity,micci2001preprocessing} \textit{Neural Network} (NN), where categorical variables with high cardinality (in this case \textit{Ion}, \textit{Cells} and \textit{CellCycle}) are pre-processed to learn a new meaningful data representation. Once the initial parameters are selected, (e.g. cell line, ion type and kinetic energy), the survival is calculated with each of the four models, and these values are these used as input to ANAKIN to predict the final cell survival. 

Tree-based models have been chosen for the predictive modules rather than \textit{Neural Network} (NN)-based algorithms because, to date, despite the \red{groundbreaking} impact that NN had on image detection, NN had a significant less impact on tabular data; there are in fact several empirical evidences that standard ML approaches have comparable or even better results than NN, \cite{grinsztajn2022tree}. On the contrary, DL algorithms are used within ANAKIN in an innovative way to solve a different task. As mentioned above, ANAKIN is trained to predict the cell survival over many cell-lines as well as ions. Such variables can assume only discrete values, typically referred to as \textit{categorical variables} in the ML and DL community, and for this reason they are not in principle easily handled by a ML or DL model. Even more problematic there is the fact that such categorical variables have a high number of possible values. This poses a serious issue in how these variables must be mapped to  \red{numeric} values to be efficiently treated by a ML model. Several possible solutions to the above problem exist, \cite{seger2018investigation}, but recently, DL have gained a huge attention not as solely a predictive tool but also as an extremely powerful data pre-processing tool, used for instance as a model to extract new information from data. For instance, DL has been recently proposed to specifically treat categorical variable with a high number of values. Such technique is called \textit{Deep Embedding}, \cite{micci2001preprocessing,guo2016entity,deep_emb}, and consists in training a NN that learns the most efficient way of encoding a categorical variable, such as in the present case the cell-line or also the ion type, into a low-dimensional numerical vector that can be efficiently used by another model to understand the most accurate relation between these variables and the target variable to predict. Therefore, ANAKIN has three specifically devoted module to learn a new data representation for the cell-lines, ion type and also cell cycle. The DL-based Deep Encoding modules are connected to the previously mentioned tree-based predictive modules to create ANAKIN, the final ensemble model that takes each single module output and predicts the cell survival fraction.

\begin{comment}
\begin{figure}
    \centering
 \includegraphics[width=.9\columnwidth]{Figure/ANAKINScheme.png}
        \caption{A schematic representation of ANAKIN. Data from PIDE are input into two types of tree-based models (RF and XGBoost), either directly or after being processed with a Deep Embedding. All four models predict cell survival, and the values are used as input for ANAKIN, that opportunely combines the predictions to provide a final cell survival output.}\label{FIG:AS}
\end{figure} 
\end{comment}

Each input model has been validated using a 10-fold cross-validation, and their hyper-parameters have been obtained using a Bayesian optimization technique, as described in \cite{missiaggia2022exploratory}.

Consider dataset $\mathcal{D}:= \left \{\mathcal{X}_i, y_i \right \}_{i=1}^N$ composed by $N$ samples, where
\[
\mathcal{X}_i:=\left (x^i_1,\dots,x^i_n\right )\,,\quad \, n \in \mathbb{N}\,, i=1,\dots,N < \infty\,,
\]
are the $n$ input features on which a model is trained to predict the target variable $y_i \in \mathbb{R}$. In the current case, $\mathcal{X}_i$ are the variables reported in Table \ref{TAB:PIDE}, whereas $y_i$ is the cell survival.

Given a set of parameter $W$, that depends on the model, and a suitably chosen training set $\mathcal{T} := \left \{\mathcal{X}_i, y_i \right \}_{i=1}^{N_{\mathcal{T}}}$, $N_{\mathcal{T}} < N$, the aim of the ML or DL models is to solve the following optimization problem
\[
\min_W \, \sum_{t=1}^N \mathcal{L}\left (\mathcal{Y}_i,\hat{\phi}\left (\mathcal{X}_i;W\right )\right )\,,
\]
being $\hat{\phi}$ the output function for the model and $\mathcal{L}$ the loss function. To improve the accuracy and reduce the overfitting, a regularization is added to the loss function, as done in \cite{bishop1995neural,lecun2015deep}.

The output function $\hat{\phi}$ is the learned function approximating the ideal function $\phi$,  that describes the link between the features $\mathcal{X}_i$ and the target $y_i$

\subsubsection{Ensemble tree-based models: Random Forest (RF) and Extreme Gradient Boosting (XGBoost)}

\textit{Random Forest} (RF) is an ensemble ML algorithm that combines weaker models, such as decision trees, to create a more robust final model \cite{ho1998random,ho1995random}. Being a bagging algorithm, the ensemble model is created in parallel, and thus the output is the average of all trees outcomes. Compared to decision-trees, the \textit{Random Forest} reduces the overfitting on the train data, and thus it improves the prediction accuracy.

RF \cite{Ho,Fri} assumes that $\hat{\phi}$ is the average of weaker learners decision--trees $\psi_k$, that is
\[
\hat{\phi}\left (x^i_1,\dots,x^i_n\right ) = \frac{1}{K} \sum_{k=1}^K \psi_k\left (x^i_1,\dots,x^i_n\right )\,,
\] 
where $\psi_k$ is the outcome of the $k-$th decision tree.

Like RF, also XGBoost is an ensemble ML algorithm that combines weaker decision trees models to create a more robust final model \cite{Chen2016,ChenG16}. XGBoost is a boosting algorithm, so that the ensemble model is created in series, and thus the output of each single model is passed to another, with the aim of reducing the error of the previous one. Also bagging is mostly used to reduce the overfitting on the train data and to improve the predictions accuracy.

XGBoost starts with a potential inaccurate model
\begin{equation}\label{EQN:Lab1}
\hat{\phi}_0(x;\bar{W}) = \arg \min_W \sum_{t=1}^N \mathcal{L}\left (\mathcal{Y}_i,\hat{\phi}\left (\mathcal{X}_i;W\right )\right )\,,
\end{equation}
and then it thus expanded in a greedy fashion as
\begin{equation}\label{EQN:Lab2}
\hat{\phi}_m(x;\bar{W}) = \hat{\phi}_{m-1}(x;\bar{W}) + \arg \min_W \left [\sum_{t=1}^N \mathcal{L}\left (\mathcal{Y}_i, \hat{\phi}_{m-1}(x;W)+\hat{\phi}\left (\mathcal{X}_i;W\right )\right )\right ]\,.
\end{equation}

\subsubsection{Deep Embedding}

Deep Embedding, \cite{micci2001preprocessing,guo2016entity,deep_emb} is a NN- based technique for mapping a categorical variable into a vector. Being a supervised algorithm, the NN is trained to predict the the cell survival fraction. Thus, the intermediate representation learnt by the network is extracted and constitutes the new values used for the categorical variable.

In the context of NN, the $W$ parameters defined in equations \eqref{EQN:Lab1}--\eqref{EQN:Lab2} are usually referred to as weights. In this work, we chose the \textit{multilayer perceptron} (MLP) NN, which is the first and most classical type of network used.

%The name of NN is taken from the similarity on how the network and the human brain work. In particular a NN is constituted of nodes, called \textit{neurons}, structured in layers, that take some input data and output the needed results.

A Multi--Layer Perceptron (MLP) is created by connecting several single layer perceptrons, where several nodes are placed in a unique layer. The inputs $\left (x_1,\dots,x_n\right )$ are fed to the network so that the final output $z$ is produced. Typically the output is a non--linear function of a weighted average of the input, i.e.
\[
y = \hat{\phi}(x) = \sigma\left (\sum_{i=1}^n w_i x_i + b\right )\,,
\]
where $w_i$ are the weights and $b$ is the bias. Also, $\sigma$ is a suitable (possibly) non--linear function, like a sigmoid 
\[
\sigma(z) = \frac{1}{1+e^{-z}}\,.
\]
%An archetype MLP is depicted in Figure \ref{FIG:mlp}.

The connection between single layer perceptrons is done in a preferred direction. This type of network is called feed--forward, because the inputs are fed to the first layer, then the output goes to the second layer and so on until the data reaches the last output layer. By providing a series of corrects results to the network, and thus making the problem supervised, the NN can learn the best weights and bias to reproduce any desired output \cite{bishop1995neural,lecun2015deep}.

\begin{comment}
\begin{figure}[thpb]
\centering
\begin{tikzpicture}[shorten >=1pt,->,draw=black!50, node distance=\layersep,
    neuron/.style={circle,fill=black!25,minimum size=17pt,inner sep=0pt},
    input neuron/.style={neuron, fill=green!40},
    output neuron/.style={neuron, fill=red!40},
    hidden neuron/.style={neuron, fill=blue!40},
    pics/graph/.style={code={\draw[double=orange,white,thick,double distance=1pt,shorten
     >=0pt]      plot[variable=\t,domain=-0.5:0.5,samples=51] 
     ({\t},{#1});}}]

    % Input layer
    \foreach \name / \y in {1,...,2}
        \node[input neuron] (I-\name) at (0,0.5-2*\y) {$i\y$};

    % Hidden layer
    \foreach \name / \y in {1,...,5}
        \path[yshift=0.5cm]
            node[hidden neuron] (H-\name) at (2.5,-\y cm) {$h\y$};

    % Output node
    \node[output neuron, right of=H-3] (O) {$o$};

    % Connect every node in the input layer with every node in the hidden layer.
    \foreach \source in {1,...,2}
        \foreach \dest in {1,...,5}
            \path (I-\source) edge (H-\dest);

    % Connect every node in the hidden layer with the output layer
    \foreach \source in {1,...,5}
        \path (H-\source) edge (O);
\end{tikzpicture}
\caption{A typical feed-forward MLP NN structure.}
\label{FIG:mlp}
\end{figure}
\end{comment}

\subsection{Explainable Artificial Intelligence}\label{SEC:MMInt}

Several XAI techniques  \cite{molnar2020interpretable,biecek2021explanatory} can be used to understand how a ML model work. in the present work, we focus on three specific very \red{well--known} and powerful techniques, namely (i) variable importance, (ii) Accumulated Local Effect (ALE) plot and (iii) the Shapley value.

\subsubsection{Variable importance}

Variable importance \cite{breiman2001random,fisher2019all} measures the global importance of each feature to the final output of the model. The main idea behind the calculation is that, if a variable is important for calculating the final output of the model, then after a permutation of the variable values, the model performance significantly decreases. Larger changes in the overall model performance are then associated to highly important features.

\subsubsection{ALE plot}

Accumulated Local Effect (ALE) plot, \cite{apley2020visualizing,gromping2020model}, is one of the most advanced and robust dependence plot for describing how variables influence on average predictions of a ML model. On of the most advanced aspects of this model is that it accounts for correlation between variables. ALE plot thus calculate the average changes in the model prediction and sum (accumulate) them over the values assumed by a specific variable.

Ale plot is defined as \cite{apley2020visualizing}
\begin{equation}\label{EQN:ALE}
\hat{f}_{\mbox{ALE}}(x_1) := \int_{x_{min}}^{x_1} \int \frac{\partial \hat{\phi}(z_1,x_2)}{\partial z_1} p(x_2|z_1) dx_2 dz_1 - \mbox{constant}\,.
\end{equation}

Instead of considering the effect of the prediction $\hat{\phi}$, the ALE plot considers changes in the prediction $\frac{\partial \hat{\phi}(z_1,x_2)}{\partial z_1}$, which represents the local effect of the variable. This is averaged over all possible values of the other variable $x_2$, weighted by the actual probability of registering the value $x_2$ given the considered value $x_1$. Then, the result is \red{integrated}, or accumulated, up to $x_1$. This value is centered around the average prediction, represented by the constant appearing in equation \eqref{EQN:ALE}, so that the average effect over the data is 0. Therefore, ALE \red{plots} calculate the average difference in the prediction to be imputed to a local \red{change} in a variable.

\subsubsection{The SHAP value}

The SHapley Additive exPlanation (SHAP) value, \cite{lundberg2017unified}, is a local XAI technique extremely powerful that aims at explaining individual predictions and in particular what is the contribution of each single variable to the overall prediction. The SHAP method computes Shapley values \cite{hart1989shapley} as an additive feature attribution, alike a linear model, so that the prediction is decomposed as
\[
\hat{\phi}(x) = \varphi_0 + \sum_{i=1}^n \varphi_i\,,
\]
where $\varphi_i$ represents the contribution of the $i-$th feature and $\varphi_0$ is an intercept.

%Among the most relevant advantages, SHAP has a solid theoretical foundation, is relatively easy to understand and it gives powerful insights into the model prediction. 
%\subsection{Software used}

\subsection{Error assessment}\label{SEC:Error}

To provide a comprehensive and accurate assessment of ANAKIN performances, many metrics are used throughout this paper. In order to compare cell survival fractions, for each \red{experiment} we computed the \textit{logarithmic Root Mean Square Error} (\textit{logRMSE}), defined as
\[
\mbox{logRMSE}^i := \sqrt{\frac{1}{N_D} \sum_{D} \left (\log \hat{S}^i(D) - \log S^i(D)\right )^2}\,,
\]
where $D$ is the dose and $N_D$ is the number of doses \red{measured} in the $i-$th experiment. $\hat{S}^i$ and $S^i$ are the cell survivals predicted and measured, respectively. In the paper, the average and standard deviation of all the errors used are calculated by averaging the results for experiments included in the test set.

The RBE at the \red{survival} level $\rho$ is defined as
\[
RBE_\rho = \frac{\sqrt{\alpha_\gamma^2-4\beta_\gamma \log \rho}-\alpha_X}{2\beta_\gamma D}\,,
\]
where $D$ is the dose giving $\rho$ survival fraction. Also, we denote
\[
RBE_\alpha = \frac{\alpha_{\mbox{ion}}}{\alpha_\gamma}\,,\quad RBE_\beta := \sqrt{\frac{\beta_{\mbox{ion}}}{\beta_\gamma}}\,,
\]
where \AG and \BG represent the $\alpha$ and $\beta$ value for the reference radiation, respectively. We specifically consider three \red{survival} levels at $\rho = 0.5,\,0.1,\,0.01$. In the paper, we focus on \RBE{0.1} predictions, as this is the main value used in radiobiology for particle therapy.

The comparison of RBE measured or calculated with ANAKIN is investigated using the Mean Absolute Error (MAE) metric
\[
MAE := \frac{1}{N} \sum_{i=1}^N \left | \bar{RBE}^i_\rho - RBE^i_\rho\right |\,,
\]
and the Mean Absolute Percentage Error (MAPE) metric
\[
MAPE := \frac{1}{N} \sum_{i=1}^N \frac{\left | \bar{RBE}^i_\rho - RBE^i_\rho\right |}{RBE^i_\rho}\,.
\]
$\bar{RBE}^i_\rho$ and $RBE^i_\rho$ represent ANAKIN values and measurements, respectively, for the endpoint $\rho=0.5,\,0.1,\,0.01\,\alpha,\,\beta$ and the $i-$th experiment. Since the range of RBE is extremely wide, the two metrics are often used together to provide a better evaluation of the performances of ANAKIN. 

We also calculated the MAE values of $\alpha$ and $\beta$ as
\[
MAE_\alpha := \frac{1}{N} \sum_{i=1}^N \left | \bar{\alpha}^i_{\mbox{ion}} - \alpha^i_{\mbox{ion}}\right |\,,\quad MAE_\beta := \frac{1}{N} \sum_{i=1}^N \left | \bar{\beta}^i_{\mbox{ion}} - \beta^i_{\mbox{ion}}\right |\,,
\]
where $\bar{\alpha}^i_{\mbox{ion}}$ and $\bar{\beta}^i_{\mbox{ion}}$ are the predicted $\alpha$ and $\beta$ values for the $i-$th experiment, whereas $\alpha^i_{\mbox{ion}}$ and $\beta^i_{\mbox{ion}}$ are the measured data. For those quantities, the MAPE values were not calculated, as the absolute value of both $\alpha$ and $\beta$ were close to 0.

%To guide the eyes, many \red{plots} are shown together with continuous lines\red{, which are} obtained via a spline smoothing method \cite{perperoglou2019review}.

\section{Results}\label{SEC:Res}

Results of the current work include a quantitative and comprehensive analysis of the comparison between ANAKIN cell survival predictions and experimental measurements available in PIDE. A wide range of possible metrics, such as RBE at different cell survival probabilities, $\alpha$ and $\beta$ predictions as well as the cell survival at different doses are presented. A detailed description of the used metrics is reported in Section \ref{SEC:Error}.

\begin{figure}
\begin{subfigure}[b]{\textwidth}
\includegraphics[width=.9\columnwidth]{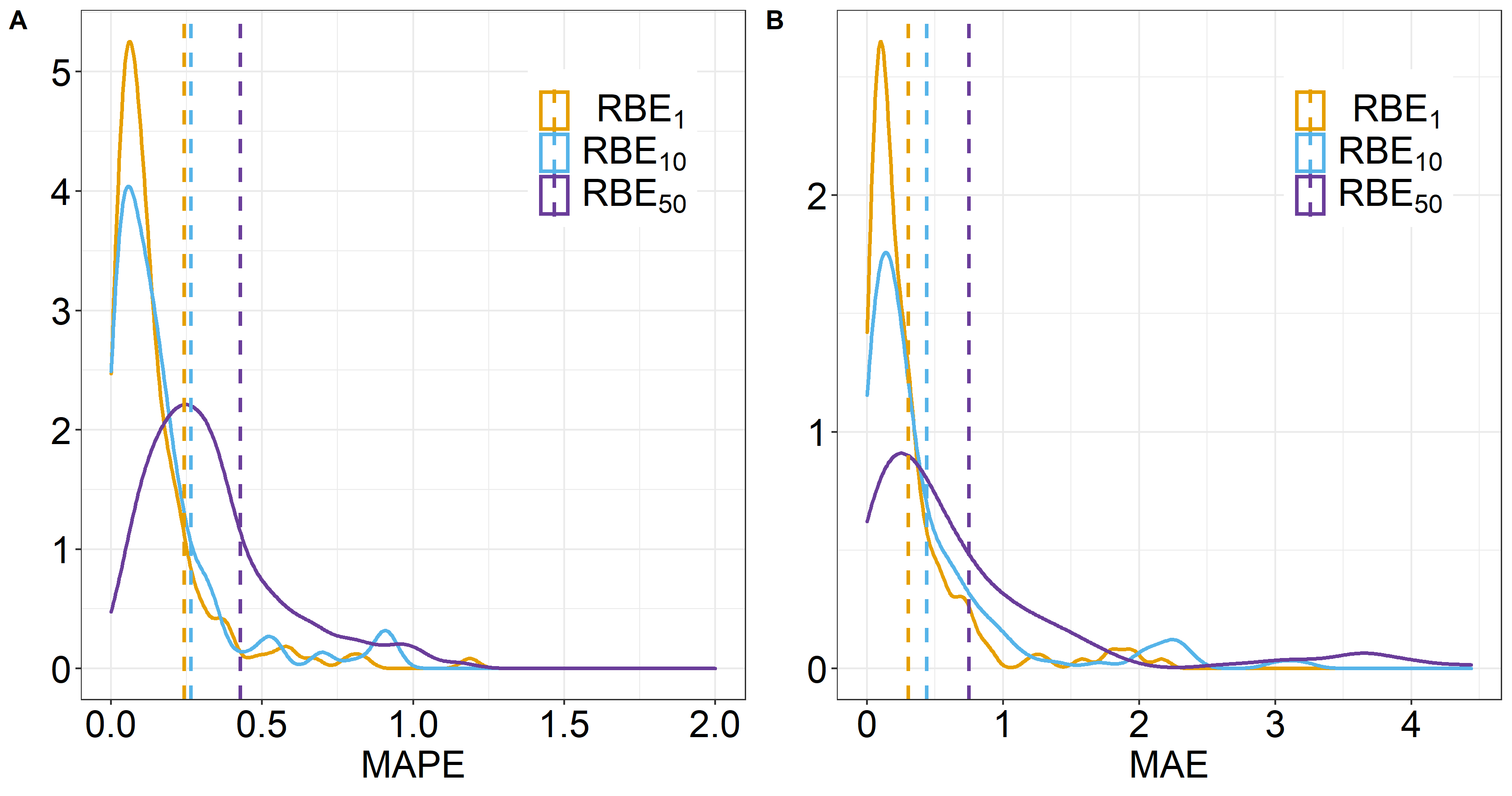}
\caption{(A) MAPE and (B) MAE distributions for \RBE{10} (yellow), \RBE{50} (blue) and \RBE{1} (purple). The dotted vertical lines indicate the average values of each distribution.}\label{FIG:2x2RBEMAE}
\end{subfigure}
\par\bigskip
\begin{subfigure}[b]{\textwidth}
\includegraphics[width=.9\columnwidth]{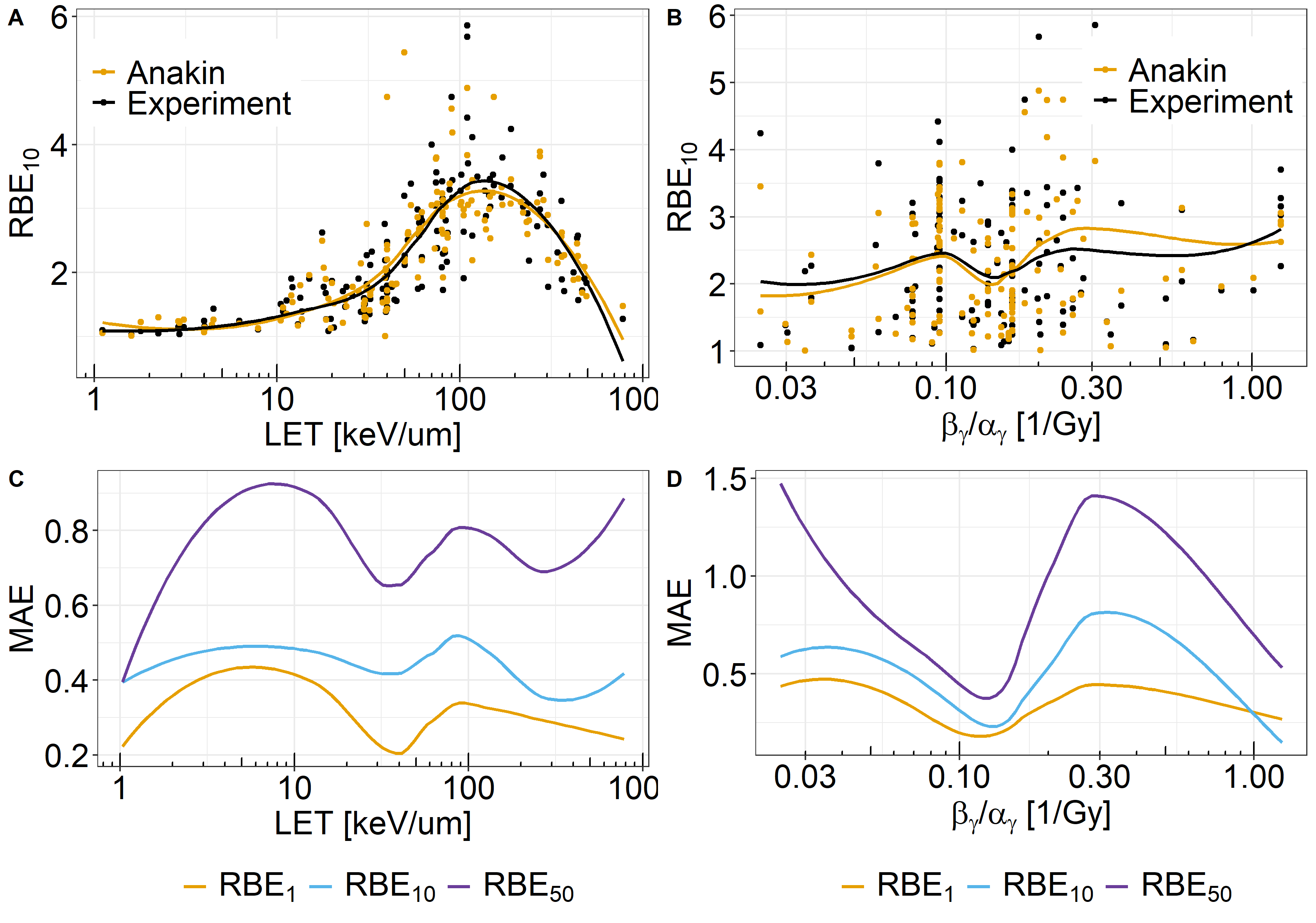}
\caption{RBE$_{10}$ predicted by ANAKIN (black) and extracted from PIDE (black) plotted against LET (A) and  $\beta_\gamma/\alpha_\gamma$ (B). To guide the eye, the continuous lines represent a spline smoothing. MAE for \RBE{10} (yellow), \RBE{50} (blue) and \RBE{1} (purple) plotted against LET (C) and  $\beta_\gamma/\alpha_\gamma$ (D).}\label{FIG:2x2RBE}
\end{subfigure}
\caption{ANAKIN predictions and error assessment for different endpoints.}\label{FIG:2x2RBEAll}
\end{figure}

Figure \ref{FIG:2x2RBEMAE} shows (A) MAPE and (B) MAE values (Section \ref{SEC:Error}) for \RBE{10}, \RBE{50} and \RBE{1}, while the numerical values as well as the \textit{logRMSE} are reported in Table \ref{TAB:Error}. The results indicate that ANAKIN has similar errors for different endpoints, with \RBE{50} exhibiting a slightly higher MAE and MAPE than \RBE{10} and \RBE{1}. 

\begin{comment}
\begin{figure}
    \centering
 \includegraphics[width=.9\columnwidth]{Figure/RBE_4x4.png}
        \caption{RBE$_{10}$ predicted by ANAKIN (black) and extracted from PIDE (black) plotted against LET (A) and  $\beta_\gamma/\alpha_\gamma$ (B). To guide the eye, the continuous lines represent a spline smoothing. MAE for \RBE{10} (yellow), \RBE{50} (blue) and \RBE{1} (purple) plotted against LET (C) and  $\beta_\gamma/\alpha_\gamma$ (D).}\label{FIG:2x2RBE}
\end{figure}
\end{comment}

\begin{table}[]
\begin{center}
\begin{tabular}{|c|l|l|l|}
\hline
\textbf{Endpoint}                          & \textbf{Error} & \textbf{Mean} & \textbf{Sd} \\ \hline \hline
                                           & logRMSE        &     1.06          &    1.26         \\ \hline\hline
\multirow{2}{*}{\RBE{10}} & MAE            &      0.43         &      0.58       \\ \cline{2-4} 
                                           & MAPE           &   0.26            &     0.80        \\ \hline
\multirow{2}{*}{\RBE{1}} & MAE            &          0.24     &        0.74     \\ \cline{2-4} 
                                           & MAPE           &   0.25           &     0.31        \\ \hline
\multirow{2}{*}{\RBE{50}} & MAE            &    0.74           &    0.91         \\ \cline{2-4} 
                                           & MAPE           &   0.42            &     0.81        \\ \hline \hline
$\alpha$ & MAE        &     0.24          &    0.24        \\ \hline
$\beta$ & MAE        &     0.03         &    0.05         \\ \hline  
\multirow{2}{*}{\RBE{\alpha}} & MAE            &          0.73     &        4     \\ \cline{2-4} 
                                           & MAPE           &   0.4           &     1.37      \\ \hline
\multirow{2}{*}{\RBE{\beta}} & MAE            &    0.43        &    0.65         \\ \cline{2-4} 
                                           & MAPE           &   0.43            &     0.59       \\ \hline                                   
\end{tabular}
\caption{Average errors and standard deviations of different error metrics and endpoints.}\label{TAB:Error}
\end{center}
\end{table}

Measured RBE$_{10}$ values and ANAKIN predictions are reported in Figure \ref{FIG:2x2RBE} as as function of the LET and $\beta_\gamma/\alpha_\gamma$. In addition, the MAE for \RBE{10}, \RBE{50} and \RBE{1} are plotted against LET and $\beta_\gamma/\alpha_\gamma$. The results indicate an excellent agreement between \RBE{10} ANAKIN and the experimental data over the entire range of LET. The smoothing spline of the \RBE{10} predicted as a function of LET completely overlaps with the experimental curve. despite this is not \red{necessarily} a proof of a perfect agreement, it is nonetheless clear that the experimental trend is predicted by ANAKIN. A good agreement can be also seen by \red{analyzing} each experiment results. This is also supported by the MAE for the other endpoints (Figure \ref{FIG:2x2RBE} (C) and Table \ref{TAB:Error}), which remains mostly constant around for LET$>$10 keV$/\mu$m. Concerning errors as a function of \AB, there is a higher variability than observed for LET. The discrepancy observed in the spline smoothing at high \BA seems an artifact of the smoothing procedure, as it is not reflected in the MAE (panel (D)). On the contrary, at low \BA, i.e. for high \AB cell--lines, ANAKIN clearly underestimates the \RBE{10}, as it is also indicated by the high MAE in the low \BA region.

\begin{figure}
    \centering
 \includegraphics[width=.9\columnwidth]{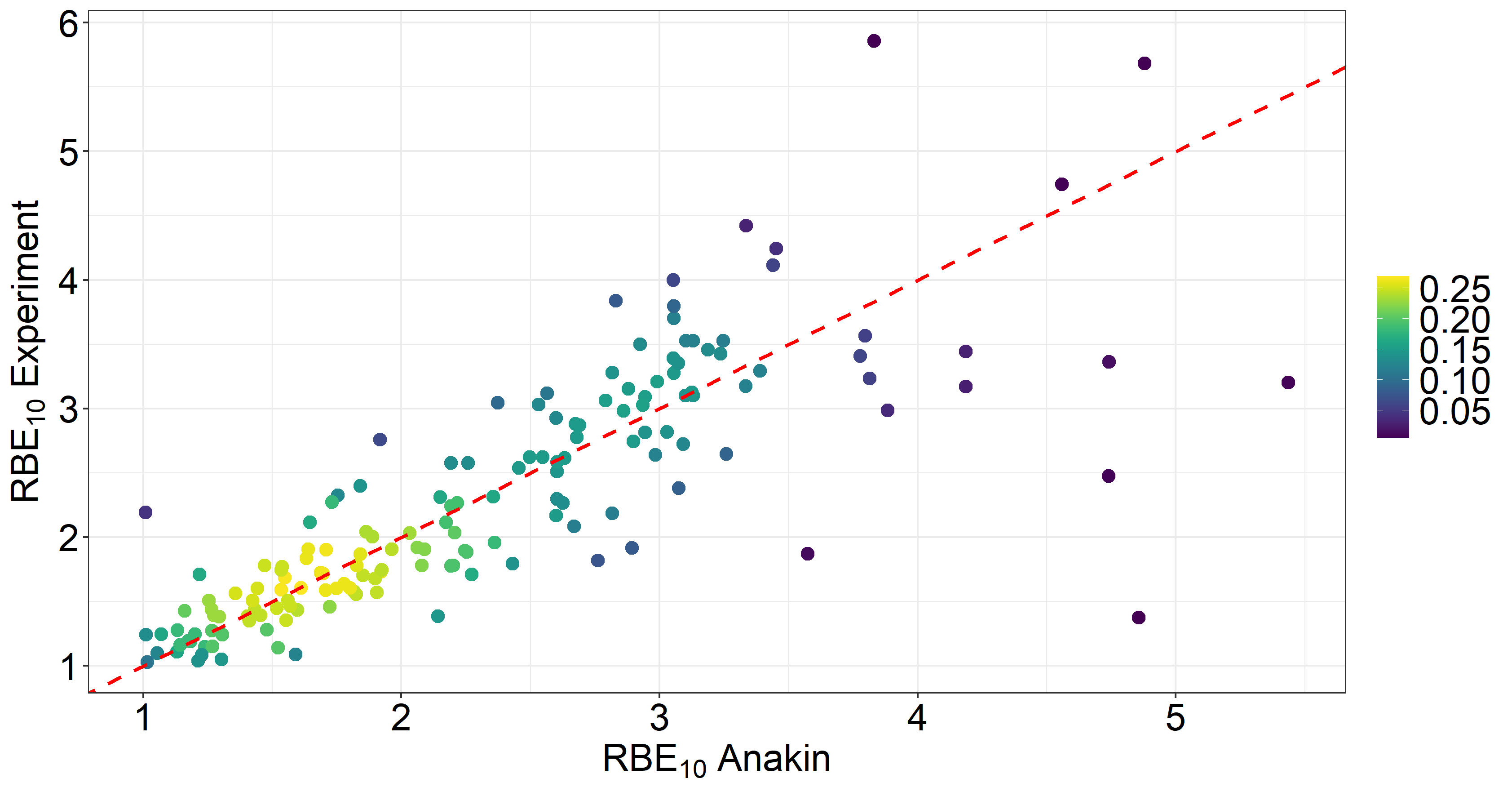}
         \caption{RBE$_{10}$ extracted from PIDE plotted against the RBE$_{10}$ predicted by ANAKIN. The color represents the density, while the diagonal dotted red line indicates the perfect prediction.}
    \label{FIG:AvsE}
\end{figure}

Figure \ref{FIG:AvsE} shows the experimental \RBE{10} against ANAKIN prediction. The results are sharply distributed around the bisector representing the ideal perfect prediction. The deviation between the bisector and the model prediction increases as the RBE grows.

\begin{figure}
    \centering
 \includegraphics[width=.9\columnwidth]{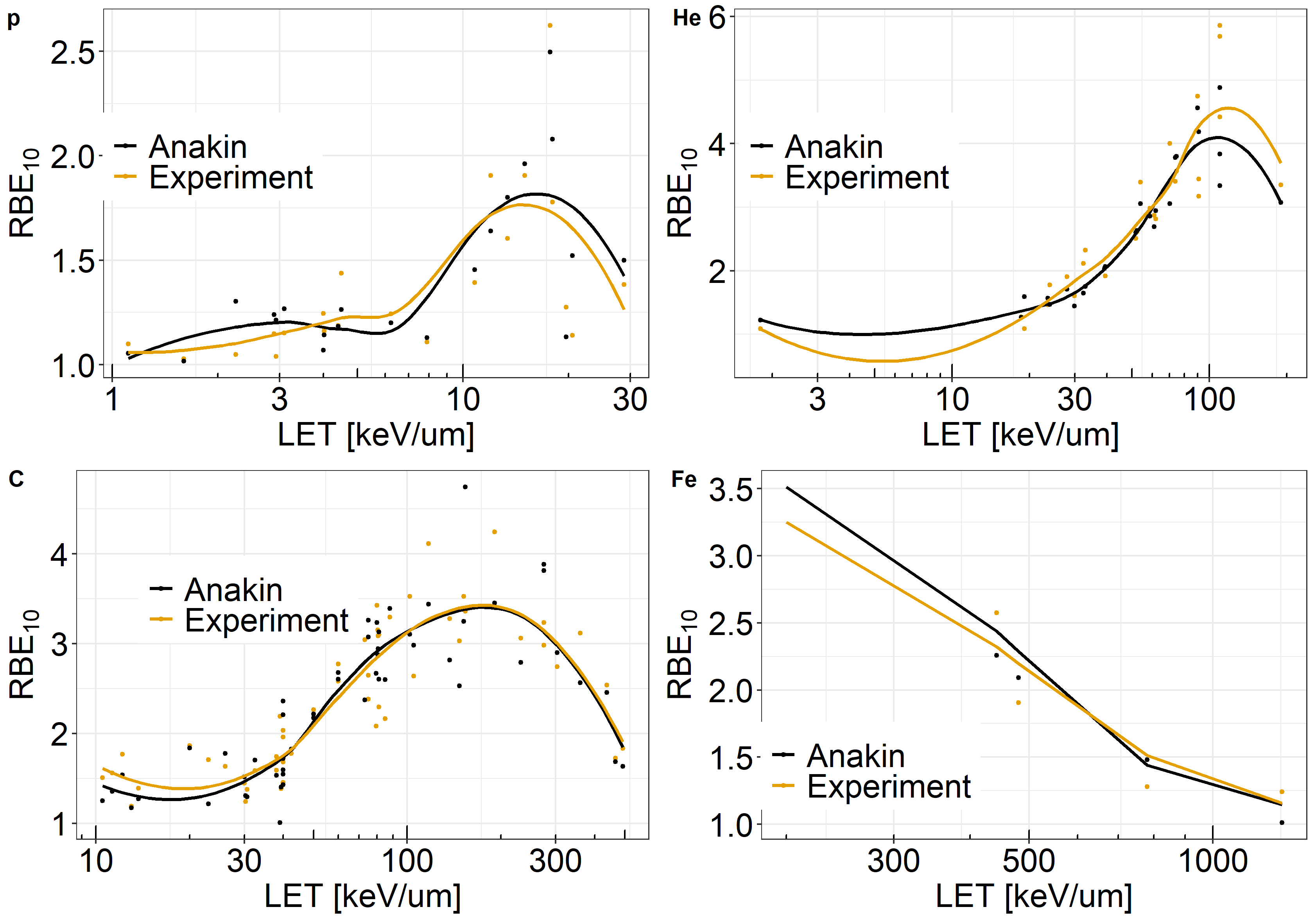}
          \caption{RBE$_{10}$ predicted by ANAKIN (black) and extracted from PIDE (yellow) plotted against LET for protons (top left), helium  (top right), carbon (bottom left) and iron(bottom right) ions. To guide the eye, the continuous lines represent spline smoothing.}\label{FIG:IonRBE}
\end{figure}

Figure \ref{FIG:IonRBE} reports ANAKIN RBE$_{10}$ predictions compared to the measurements, plotted against LET for 4 different ions (protons, helium, carbon and iron) in a very broad LET range. Overall, ANAKIN seems to reproduce well the the experimental data. For protons, ANAKIN can reproduce the small RBE variability at low LET as well as the clear rise above 20 keV$/\mu$m. ANAKIN is accurate also for helium and carbon ions, and it is clearly able to reproduce the overkilling effect, that yields a decrease in the \RBE{10} around 100 keV$/\mu$m. ANAKIN values appear to be very close to the measurements also for iron.

\begin{figure}
\begin{subfigure}[b]{\textwidth}
\includegraphics[width=.9\columnwidth]{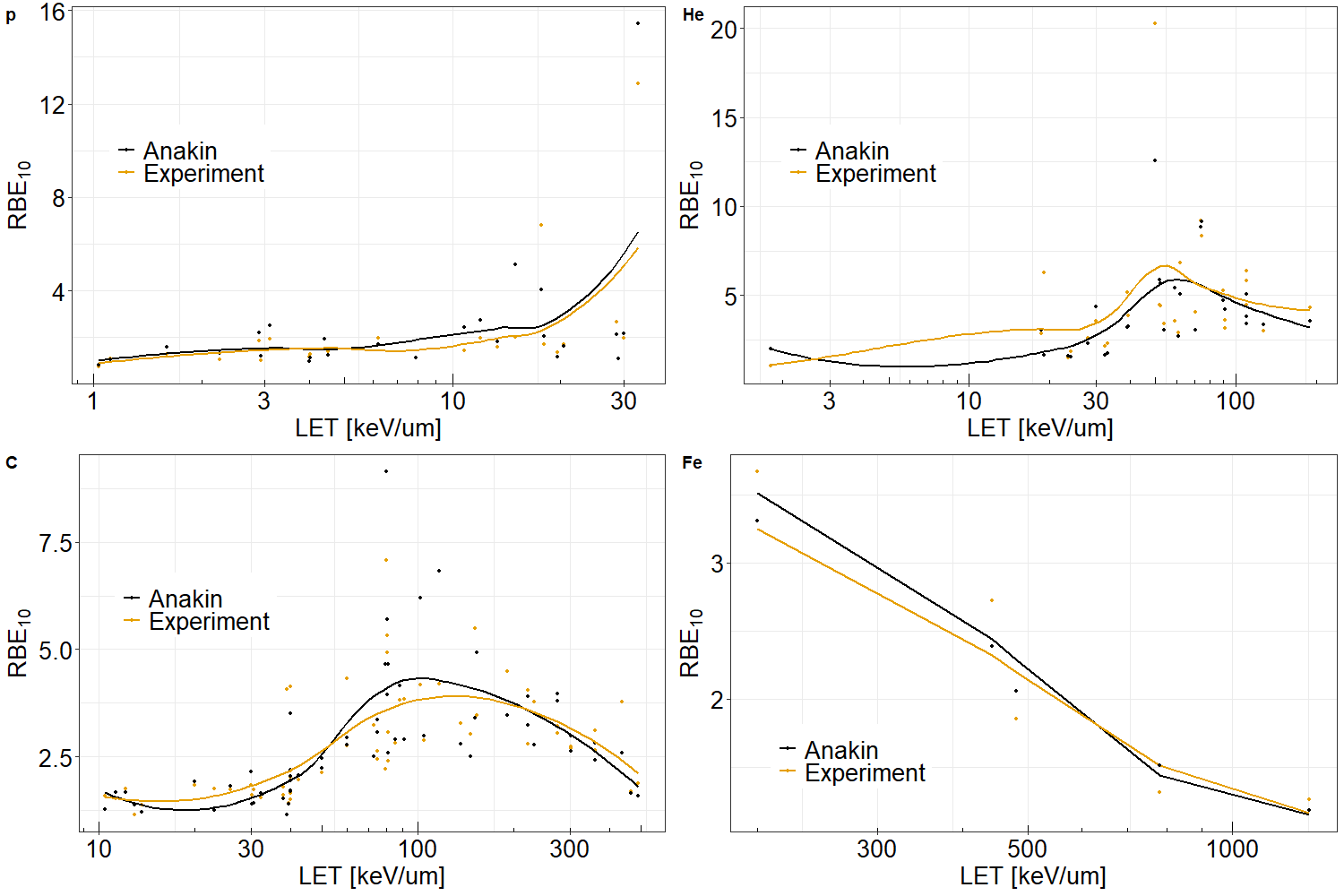}
          \caption{MAPE (A) and MAE (B) distributions for ANAKIN RBE$_{10}$ prediction for carbon ions (yellow), iron (blue), helium (purple) and protons (red). Dotted vertical lines indicate the distributions average values.}\label{FIG:IonRBEMAE}
\end{subfigure}
\par\bigskip
\begin{subfigure}[b]{\textwidth}
\includegraphics[width=.9\columnwidth]{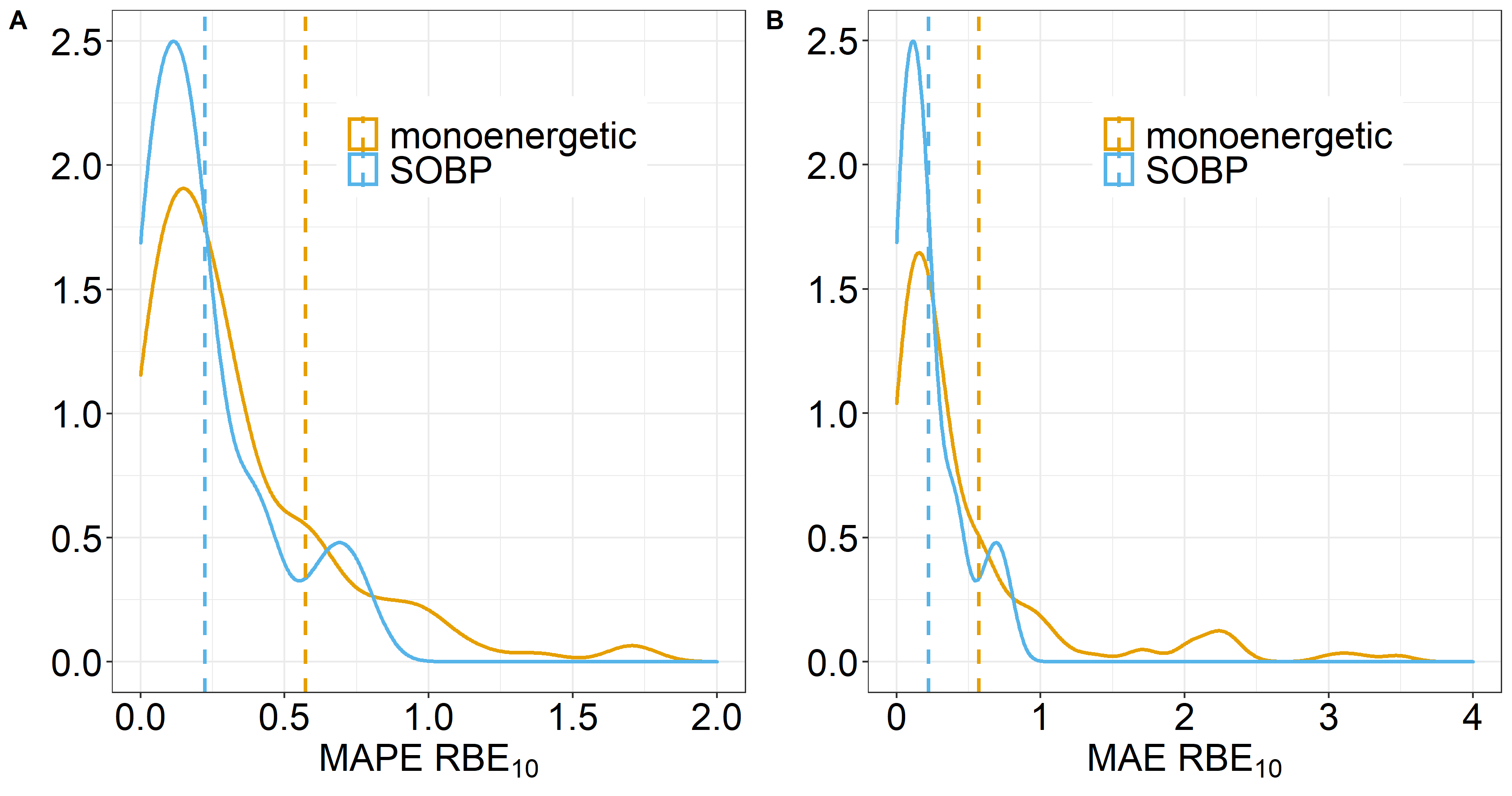}
         \caption{MAPE (A) and MAE (B) distribution for ANAKIN RBE$_{10}$ prediction for monoenergetic beams (yellow) and SOBP (blue). The dotted vertical lines indicate the distributions average values.}\label{FIG:MAPESOBP}
\end{subfigure}
\caption{ANAKIN MAPE and MAE distributions for different ions and irradiation conditions.}\label{FIG:MAPESOBPAll}
\end{figure}

\begin{comment}
\begin{figure}
    \centering
 \includegraphics[width=.9\columnwidth]{Figure/RBE_Ion.png}
          \caption{MAPE (A) and MAE (B) distributions for ANAKIN RBE$_{10}$ prediction for carbon ions (yellow), iron (blue), helium (purple) and protons (red). Dotted vertical lines indicate the distributions average values.}\label{FIG:IonRBEMAE}
\end{figure}
\end{comment}

Similar conclusions can be drawn from Figure \ref{FIG:IonRBEMAE}. Iron shows a strongly peaked distribution because of the low number of available experiments; nonetheless, iron a low error in both metrics. Beside iron, the other ions show comparable results, with helium having a broader distribution in both MAE and MAPE reflecting a lower accuracy of ANAKIN. Protons exhibit an error distribution peaked around the average values, as well as some outliers with high errors, as clearly indicated by the spikes in the high errors region. However, these peaks are for MAPE, and we hypothesize that they might be mainly caused by low RBE values, that can result in high percentage errors.

\begin{comment}
\begin{figure}
    \centering
 \includegraphics[width=.9\columnwidth]{Figure/MAPE_SOBP.png}
         \caption{MAPE (A) and MAE (B) distribution for ANAKIN RBE$_{10}$ prediction for monoenergetic beams (yellow) and SOBP (blue). The dotted vertical lines indicate the distributions average values.}\label{FIG:MAPESOBP}
\end{figure}
\end{comment}

Figure \ref{FIG:MAPESOBP} shows MAE and MAE error distributions evaluated for \RBE{10}, grouped by monoenergetic beam and Spread-out Bragg-peak (SOBP). The peak of the distributions is similar for both cases, but the error distribution for the monoenergetic beams is clearly broader than for the SOBP.

\begin{figure}
    \centering
 \includegraphics[width=.9\columnwidth]{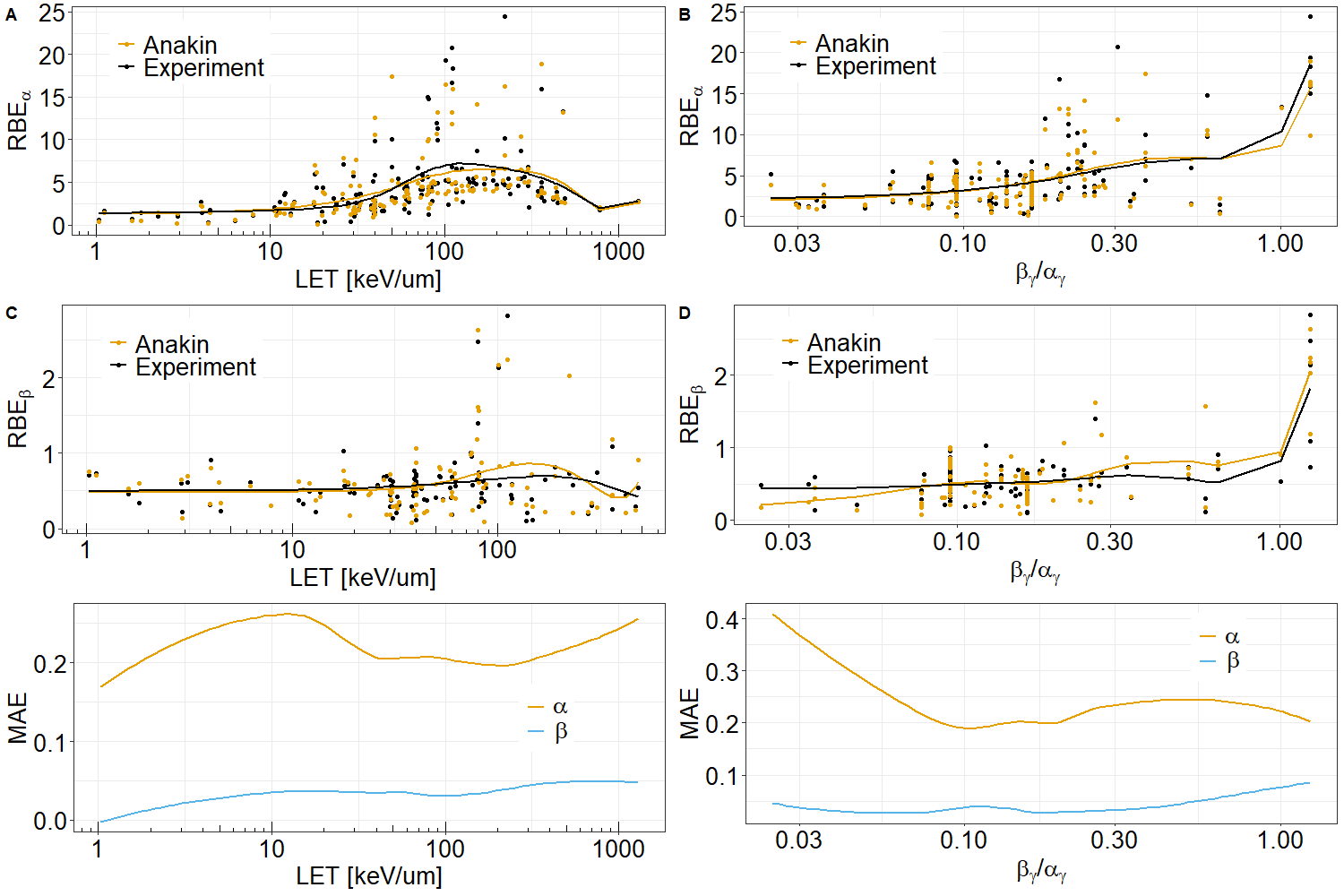}
         \caption{\RBE{\alpha} predicted by ANAKIN (yellow) and extracted from PIDE (black) plotted against LET (A) and $\beta_\gamma/\alpha_\gamma$ (B). \RBE{\beta} predicted by ANAKIN (yellow) and extracted from PIDE (black) plotted against LET (C) and $\beta_\gamma/\alpha_\gamma$ (D). The continuous lines represent spline smoothing. MAE of ANAKIN predictions and data from PIDE for $\alpha_{\mbox{ion}}$ (yellow) $\beta_{\mbox{ion}}$ (blue) values plotted against LET (E) and \BA (F).}
    \label{FIG:RBEab}
\end{figure}

Figure \ref{FIG:RBEab} shows \RBE{\alpha} and \RBE{\beta} plotted against LET and \BA. \RBE{\alpha} values are accurately predicted by ANAKIN independently of LET and \BA. A higher inaccuracy is observed for \RBE{\beta} in the low \BA region. The absolute errors in the $\alpha$ and $\beta$ predictions show a steady behavior over the LET range (panel (E)), while the errors on the $\alpha$ values clearly decreases as \BA increases, coherently with previous analysis performed above. 

%\begin{figure}
%    \centering
%  \hspace{0.7cm} $\alpha$ \hspace{5cm} $\beta$ \hfill\\
% \includegraphics[width=.49\columnwidth]{Figure/Ale_a.png}
% \includegraphics[width=.49\columnwidth]{Figure/Ale_b.png}
%         \caption{}
%    \label{FIG:ALEab}
%\end{figure}

\subsection{Comparison with MKM and LEM}\label{SEC:RMKM}
To further assess the accuracy of ANAKIN in predicting cell survival and RBE, we compared it with the only two RBE models that are currently used in clinical practice, namely the MKM  \cite{hawkins1994statistical,inaniwa2010treatment,inaniwa2018adaptation,bellinzona2021linking} and LEM \cite{kramer2000treatment,elsasser2007cluster,elsasser2008accuracy,pfuhl2022comprehensive}. To calculated the biological outcomes from the MKM and LEM III, we used the \textit{survival toolkit} \cite{manganaro2018survival}. We performed the comparison for the HSG and V79 cell-lines, because they are among the most used in radiobiological experiments, and several datasets are available in literature. For the V79 cell-lines, we used 41 different experiments with proton, helium and carbon ions, while for the HSG cell-line, we included 15 experiments conducted with helium and carbon ions. To compare the models, the same metrics introduced in Section \ref{SEC:Error} are used.

Predictions with MKM and LEM have performed with the \textit{survival toolkit} \cite{manganaro2018survival,survival} \red{including the implementation of a limited number of versions for the latter models. In particular, a} newer version of the LEM, namely the LEM IV \cite{elsasser2010quantification}, has been recently developed but it has not been used in the currently study since a freely usable version is not available. For the LEM, we used version III, as the latest version (IV) is not available. However, an extensive quantitative study has been published \cite{pfuhl2022comprehensive}, and so further quantitative comparison between the LEM IV accuracy with ANAKIN can be conducted. 

\begin{table}[]
\begin{center}
\begin{tabular}{|cc|cccccl|}
\hline
\multicolumn{2}{|c|}{\multirow{2}{*}{\textbf{}}}                             & \multicolumn{6}{c|}{\textbf{V79}}                                                                                                                                                                \\ \cline{3-8} 
\multicolumn{2}{|c|}{}                                                       & \multicolumn{2}{c|}{\textbf{ANAKIN}}                                  & \multicolumn{2}{c|}{\textbf{MKM}}                                     & \multicolumn{2}{c|}{\textbf{LEM III}}            \\ \hline
\multicolumn{1}{|c|}{\textbf{Endpoint}}          & \textbf{Error}            & \multicolumn{1}{c|}{\textbf{Mean}} & \multicolumn{1}{c|}{\textbf{Sd}} & \multicolumn{1}{c|}{\textbf{Mean}} & \multicolumn{1}{c|}{\textbf{Sd}} & \multicolumn{1}{c|}{\textbf{Mean}} & \textbf{Sd} \\ \hline
\multicolumn{1}{|c|}{}                           & logRMSE                   & \multicolumn{1}{c|}{0.70}          & \multicolumn{1}{c|}{0.71}        & \multicolumn{1}{c|}{1.9}           & \multicolumn{1}{c|}{1.43}        & \multicolumn{1}{c|}{1.66}          & 1.4         \\ \hline
\multicolumn{1}{|c|}{\multirow{2}{*}{\RBE{10}}} & MAE                       & \multicolumn{1}{c|}{0.44}          & \multicolumn{1}{c|}{0.74}        & \multicolumn{1}{c|}{1.2}             & \multicolumn{1}{c|}{0.79}        & \multicolumn{1}{c|}{1.5}           & 0.9         \\ \cline{2-8} 
\multicolumn{1}{|c|}{}                           & MAPE                      & \multicolumn{1}{c|}{0.23}          & \multicolumn{1}{c|}{0.16}        & \multicolumn{1}{c|}{0.61}          & \multicolumn{1}{c|}{0.2}         & \multicolumn{1}{c|}{0.73}          & 0.16        \\ \hline
\multicolumn{1}{|c|}{\multirow{2}{*}{\RBE{1}}}  & MAE                       & \multicolumn{1}{c|}{0.57}          & \multicolumn{1}{c|}{0.85}        & \multicolumn{1}{c|}{1.4}           & \multicolumn{1}{c|}{1.62}        & \multicolumn{1}{c|}{1.71}          & 1.7         \\ \cline{2-8} 
\multicolumn{1}{|c|}{}                           & MAPE                      & \multicolumn{1}{c|}{0.2}           & \multicolumn{1}{c|}{0.14}        & \multicolumn{1}{c|}{0.42}          & \multicolumn{1}{c|}{0.43}        & \multicolumn{1}{c|}{0.46}          & 0.47        \\ \hline
\multicolumn{1}{|c|}{\multirow{2}{*}{\RBE{50}}} & MAE                       & \multicolumn{1}{c|}{0.48}          & \multicolumn{1}{c|}{0.39}        & \multicolumn{1}{c|}{1.27}          & \multicolumn{1}{c|}{1.68}        & \multicolumn{1}{c|}{1.71}          & 2.37        \\ \cline{2-8} 
\multicolumn{1}{|c|}{}                           & \multicolumn{1}{l|}{MAPE} & \multicolumn{1}{c|}{0.17}          & \multicolumn{1}{c|}{0.06}        & \multicolumn{1}{c|}{0.39}          & \multicolumn{1}{c|}{0.36}        & \multicolumn{1}{c|}{0.4}           & 0.41        \\ \hline
\end{tabular}
\begin{tabular}{|cc|cccccl|}
\hline
\multicolumn{2}{|c|}{\multirow{2}{*}{\textbf{}}}                                             & \multicolumn{6}{c|}{\textbf{HSG}}                                                                                                                                                                \\ \cline{3-8} 
\multicolumn{2}{|c|}{}                                                                       & \multicolumn{2}{c|}{\textbf{ANAKIN}}                                  & \multicolumn{2}{c|}{\textbf{MKM}}                                     & \multicolumn{2}{c|}{\textbf{LEM III}}            \\ \hline
\multicolumn{1}{|c|}{\textbf{Endpoint}}                          & \textbf{Error}            & \multicolumn{1}{c|}{\textbf{Mean}} & \multicolumn{1}{c|}{\textbf{Sd}} & \multicolumn{1}{c|}{\textbf{Mean}} & \multicolumn{1}{c|}{\textbf{Sd}} & \multicolumn{1}{c|}{\textbf{Mean}} & \textbf{Sd} \\ \hline
\multicolumn{1}{|c|}{}                                           & logRMSE                   & \multicolumn{1}{c|}{0.72}          & \multicolumn{1}{c|}{0.42}        & \multicolumn{1}{c|}{1.06}          & \multicolumn{1}{c|}{0.47}        & \multicolumn{1}{c|}{1.36}          & 0.93        \\ \hline
\multicolumn{1}{|c|}{\multirow{2}{*}{\RBE{10}}} & MAE                       & \multicolumn{1}{c|}{0.43}          & \multicolumn{1}{c|}{0.3}         & \multicolumn{1}{c|}{1.28}          & \multicolumn{1}{c|}{0.38}        & \multicolumn{1}{c|}{1.55}          & 0.26        \\ \cline{2-8} 
\multicolumn{1}{|c|}{}                                           & MAPE                      & \multicolumn{1}{c|}{0.11}          & \multicolumn{1}{c|}{0.07}        & \multicolumn{1}{c|}{0.16}          & \multicolumn{1}{c|}{0.09}        & \multicolumn{1}{c|}{0.21}          & 0.1         \\ \hline
\multicolumn{1}{|c|}{\multirow{2}{*}{\RBE{1}}}  & MAE                       & \multicolumn{1}{c|}{0.31}          & \multicolumn{1}{c|}{0.2}         & \multicolumn{1}{c|}{1.44}          & \multicolumn{1}{c|}{0.22}        & \multicolumn{1}{c|}{1.45}          & 0.31        \\ \cline{2-8} 
\multicolumn{1}{|c|}{}                                           & MAPE                      & \multicolumn{1}{c|}{0.14}          & \multicolumn{1}{c|}{0.06}        & \multicolumn{1}{c|}{0.21}          & \multicolumn{1}{c|}{0.07}        & \multicolumn{1}{c|}{0.21}          & 0.14        \\ \hline
\multicolumn{1}{|c|}{\multirow{2}{*}{\RBE{50}}} & MAE                       & \multicolumn{1}{c|}{0.51}          & \multicolumn{1}{c|}{0.49}        & \multicolumn{1}{c|}{1.63}          & \multicolumn{1}{c|}{0.56}        & \multicolumn{1}{c|}{1.69}          & 0.34        \\ \cline{2-8} 
\multicolumn{1}{|c|}{}                                           & \multicolumn{1}{l|}{MAPE} & \multicolumn{1}{c|}{0.15}          & \multicolumn{1}{c|}{0.1}         & \multicolumn{1}{c|}{0.19}          & \multicolumn{1}{c|}{0.11}        & \multicolumn{1}{c|}{0.28}          & 0.08        \\ \hline
\end{tabular}
\end{center}
\caption{Average errors and standard deviations of ANAKIN, MKM and LEM III calculations for V79 and HSG cell-lines considering different endpoints.}\label{TAB:MKMLEM}
\end{table}

\begin{figure}
\begin{subfigure}[b]{\textwidth}
 \includegraphics[width=.9\columnwidth]{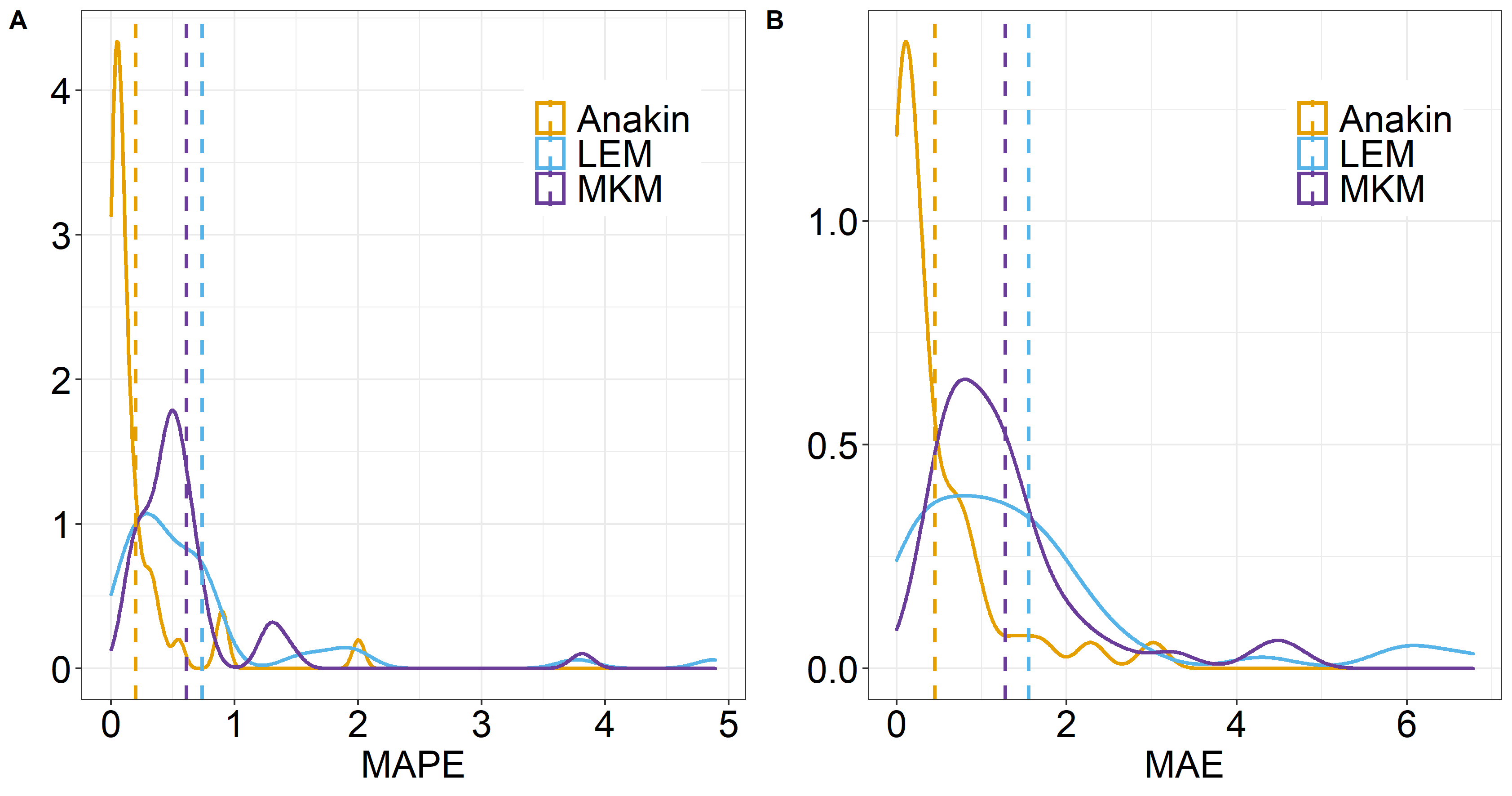}
       \caption{MAPE (A) and MAE (B) distributions for RBE$_{10}$ calculated with ANAKIN (yellow), LEM III (blue) and MKM (purple) for the V79 cell line. The dotted vertical lines denoted the average values.}\label{FIG:V79}
\end{subfigure}
\par\bigskip
\begin{subfigure}[b]{\textwidth}
 \includegraphics[width=.9\columnwidth]{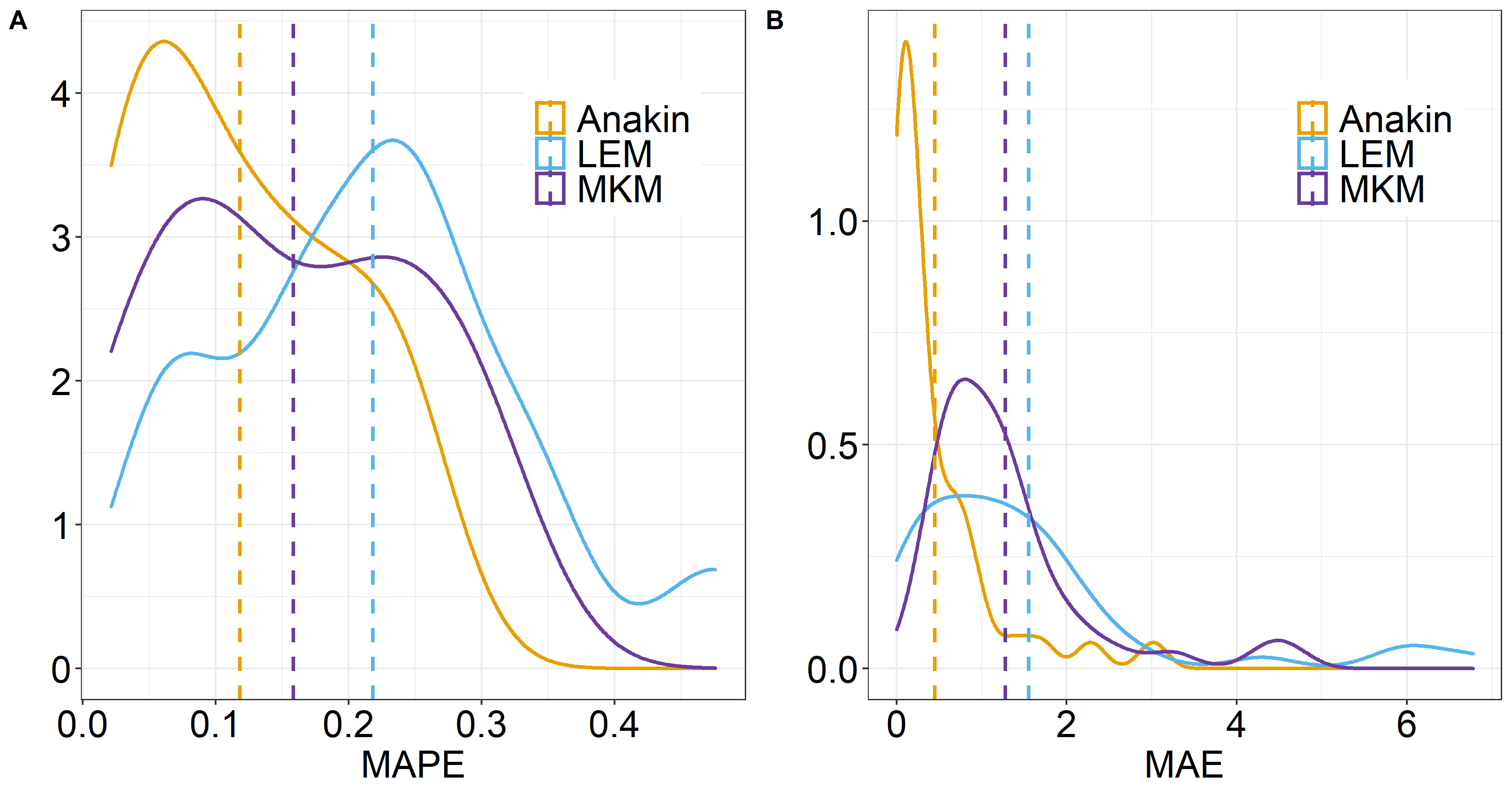}
\caption{MAPE (A) and MAE (B) distributions for RBE$_{10}$ calculated with ANAKIN (yellow), LEM III (blue) and MKM (purple) for the HGS cell line. The dotted vertical lines denoted the average values}\label{FIG:HSG}
\end{subfigure}
\caption{ANAKIN, MKM and LEM MAE and MAPE distribution for V79 and HSG cell-lines.}\label{FIG:V79All}
\end{figure}

The comparison between the models is shown in Figures \ref{FIG:V79}--\ref{FIG:HSG}, while the numerical values are reported in Table \ref{TAB:MKMLEM}. The results indicate that overall ANAKIN is more accurate than both the MKM and the LEM in predicting all metrics and endpoints considered. For both cell-lines, the LEM shows the largest deviations from the measurements, closely followed by the MKM with ANAKIN reporting lower errors. In particular, for HSG cell-line, the LEM shows a MAE for the \RBE{10} of 1.55, whilst the MKM and ANAKIN has respectively a MAE 1.18 and 0.43. For the V79 cell-line, the MKM and the LEM predict comparable results with a MAE for \RBE{10} of 1.5 and 1.2. ANAKIN shows a MAE for \RBE{10} of 0.43. Other endpoints and metrics has comparable results. Together with having lower average errors, ANAKIN exhibits a narrower error distributions, and does never reach absolute errors as high as the MKM and LEM.

\begin{comment}
\begin{figure}
    \centering
 \includegraphics[width=.9\columnwidth]{Figure/V79.png}
       \caption{MAPE (A) and MAE (B) distributions for RBE$_{10}$ calculated with ANAKIN (yellow), LEM III (blue) and MKM (purple) for the V79 cell line. The dotted vertical lines denoted the average values.}\label{FIG:V79}
\end{figure}

\begin{figure}
    \centering
 \includegraphics[width=.9\columnwidth]{Figure/HSG.png}
\caption{MAPE (A) and MAE (B) distributions for RBE$_{10}$ calculated with ANAKIN (yellow), LEM III (blue) and MKM (purple) for the HGS cell line. The dotted vertical lines denoted the average values}\label{FIG:HSG}
\end{figure}
\end{comment}

%Figures \ref{FIG:V79}--\ref{FIG:HSG} gather results contained in Table \ref{TAB:MKMLEM}, showing MAE and MAPE error distributions for \RBE{10}, \RBE{1} and \RBE{50} for V79 and HSG cell-lines, respectively. Analogous conclusions as above can be drawn, with ANAKIN showing clearly lower errors in both \red{metrics} and both cell-lines, followed \red{respectively} by the MKM and \red{the} LEM. It can be also seen how, besides having lower average errors, ANAKIN \red{presents} the less broad error distributions and, \red{in addition, it}  never exhibits \red{larege absolute errors} as instead emerging in both MKM and LEM \red{predictions}.

\begin{figure}
    \centering
 \includegraphics[width=.9\columnwidth]{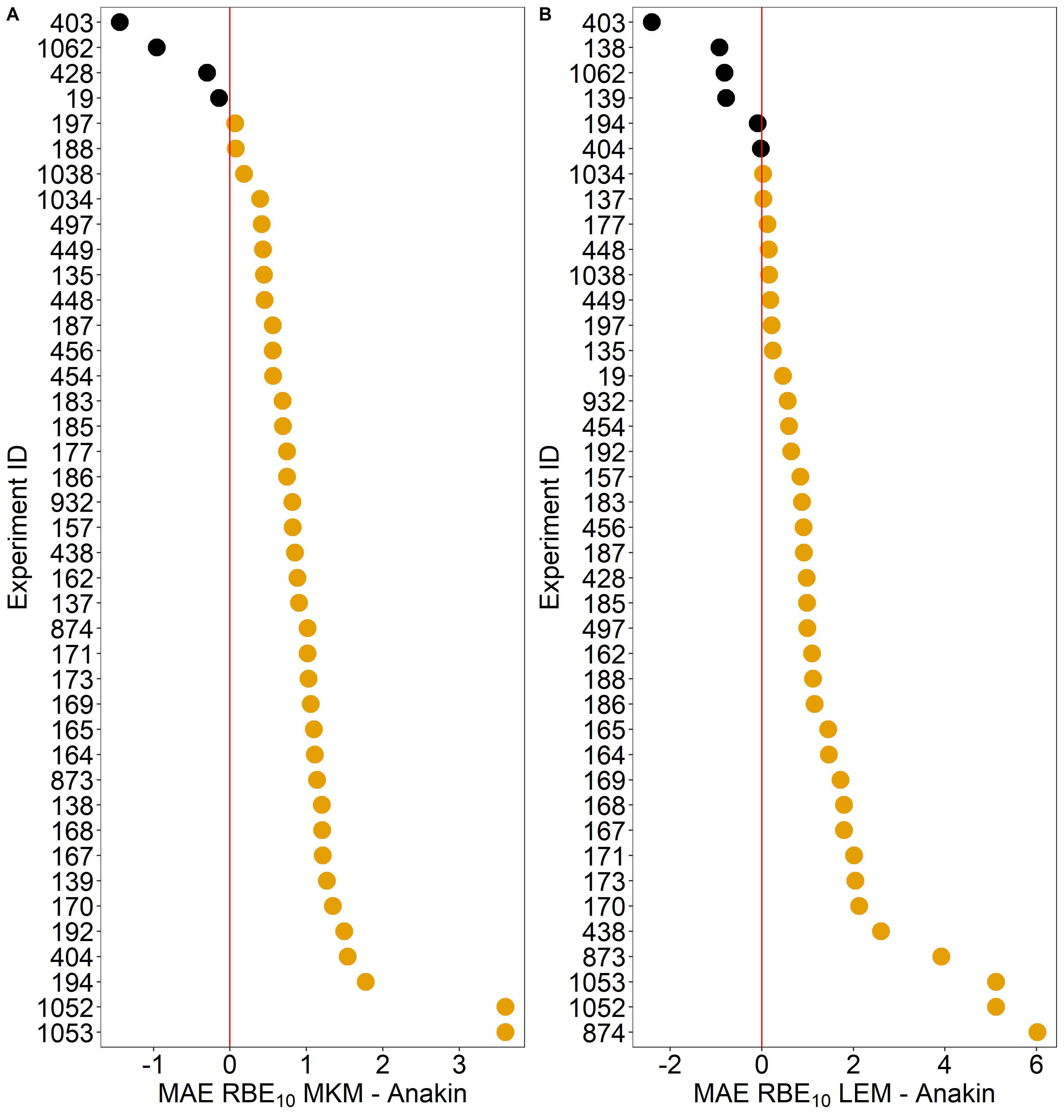}
         \caption{%MAE comparison between ANAKIN and MKM (A), and LEM (B) for the V79 cell line. Plot shows difference in MAPE for RBE$_{10}$ between (A) MKM and (B) LEM and ANAKIN***.
         MAE difference between ANAKIN and MKM (A) or LEM III (B) calculated for different experiments (labeled with an ID number on the $y$ axis). All values are for the RBE$_{10}$ of the V79 cell line available in PIDE. The black dots represents measurements for which the MKM or LEM exhibit lower MAE than ANAKIN, while the yellow dot those for which ANAKIN MAE is lower. The red vertical line indicates the zero.}\label{FIG:V79MAPE}
\end{figure}

\begin{figure}
    \centering
 \includegraphics[width=.9\columnwidth]{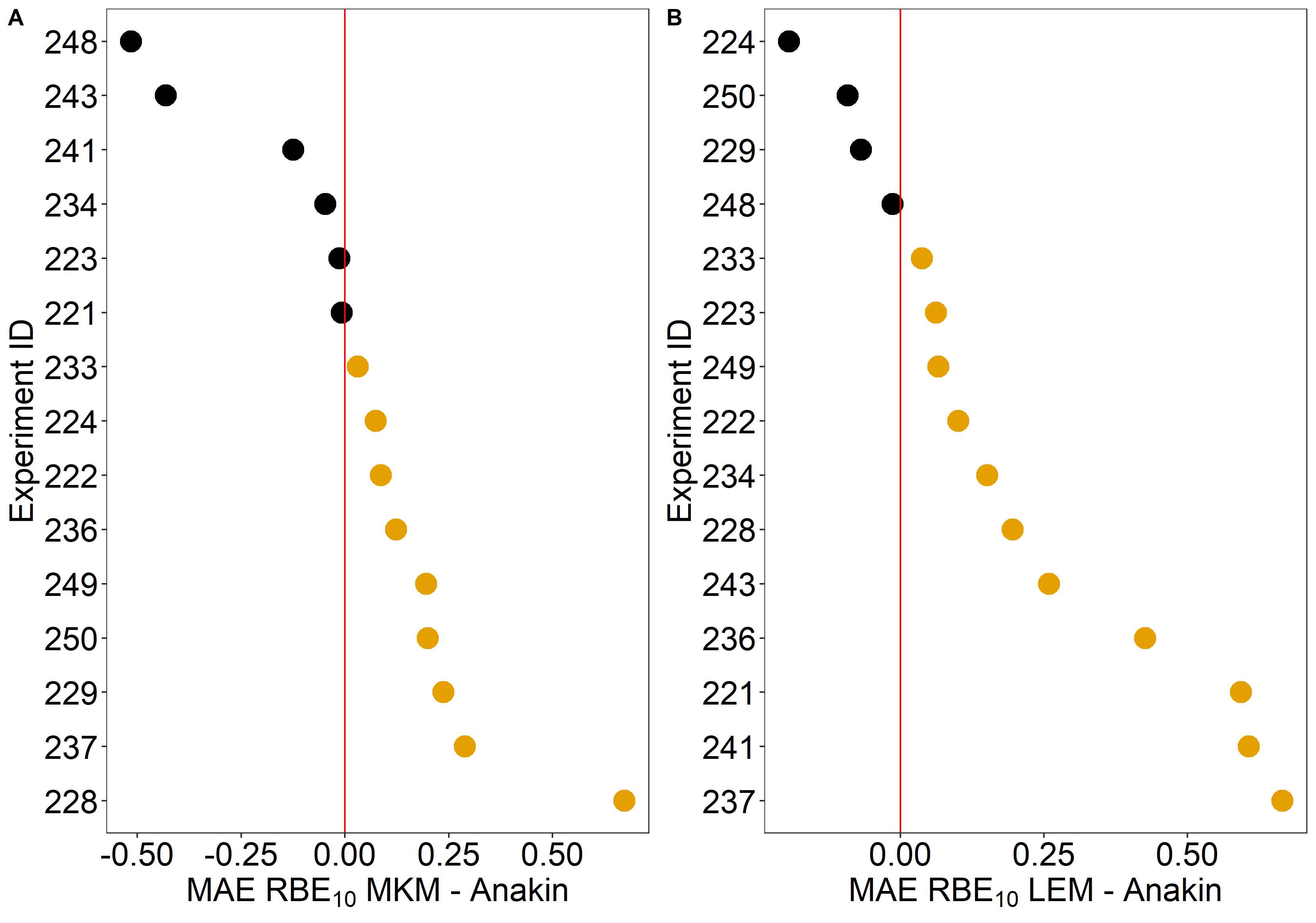}
        \caption{%MAE comparison for HSG cell line between ANAKIN and (A) MKM \red{or} (B) LEM. Plot shows difference in MAPE for RBE$_{10}$ between (A) MKM and (B) LEM and ANAKIN. Each horizontal line corresponds to a different experiment. Black values represents experiments for which (A) MKM and (B) LEM exhibits lower MAPE than ANAKIN, yellow values on the contrary represent experiments for which ANAKIN MAPE is lower. The red vertical line shows 0.
        MAE difference between ANAKIN and MKM (A) or LEM III (B) calculated for different experiments (labeled with an ID number on the $y$ axis). All values are for the RBE$_{10}$ of the HGS cell line available in PIDE. The black dots represents measurements for which the MKM or LEM exhibit lower MAE than ANAKIN, while the yellow dot those for which ANAKIN MAE is lower. The red vertical line indicates the zero.}\label{FIG:HSGMAPE}
\end{figure}

Figures \ref{FIG:V79MAPE}--\ref{FIG:HSGMAPE} illustrates the absolute difference of MAE between ANAKIN and MKM or LEM, calculated for the \RBE{10} of both cell lines. All experimental datasets were obtained from PIDE.
%The value in the horizontal axis is the absolute difference in the MAE for \RBE{10} of the MKM or LEM and ANAKIN. Black dots highlight experiments in which either the MKM or LEM has lower error compared to ANAKIN, whereas yellow dots on the opposite denote experiments where ANAKIN has lower error than the MKM or LEM.

This analysis confirms the results shown in Figures \ref{FIG:V79}--\ref{FIG:HSG} and Table \ref{TAB:MKMLEM}. ANAKIN is more accurate in predicting the selected biological outputs than both the MKM and LEM for the majority of experiments. The maximum discrepancy between MAE is significantly higher when ANAKIN is closer to the experimental data (yellow dots), reaching differences above 6 for V79 cell-lines, but only 2 for the opposite case. This result indicate that for the cases where ANAKIN is less accurate than the other models, its error is smaller than when the predictions of the MKM and LEM III are off.

%It is worth stressing that,  the MKM is particularly accurate in predicting RBE for the HSG cell-line. 

\subsection{Explainable Artificial Intelligence}\label{SEC:RXAI}

%The presented section is devoted to ANAKIN \red{results} explainability and interpretability. 

\begin{figure}
\begin{subfigure}[b]{\textwidth}
    \includegraphics[width=.9\textwidth]{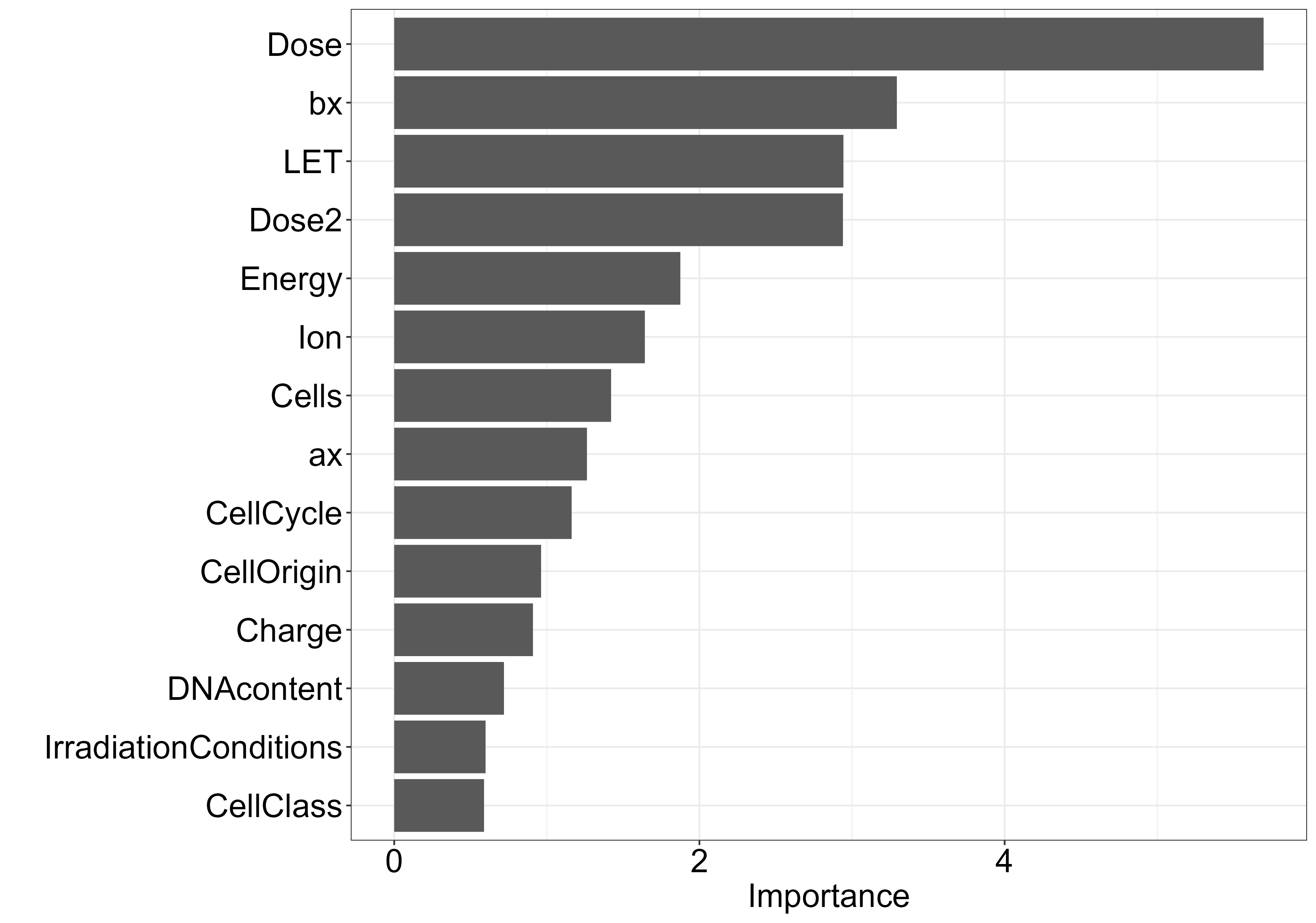}
    \caption{Importance \red{analysis} for the variables used by ANAKIN. The variable names reflect PIDE documentation \cite{friedrich2013systematic,friedrich2021update}. We also introduced the Dose2 as a new variable, representing the square of the dose.}\label{FIG:VarImp}
\end{subfigure}
\par\bigskip
\begin{subfigure}[b]{\textwidth}
    \includegraphics[width=.9\textwidth]{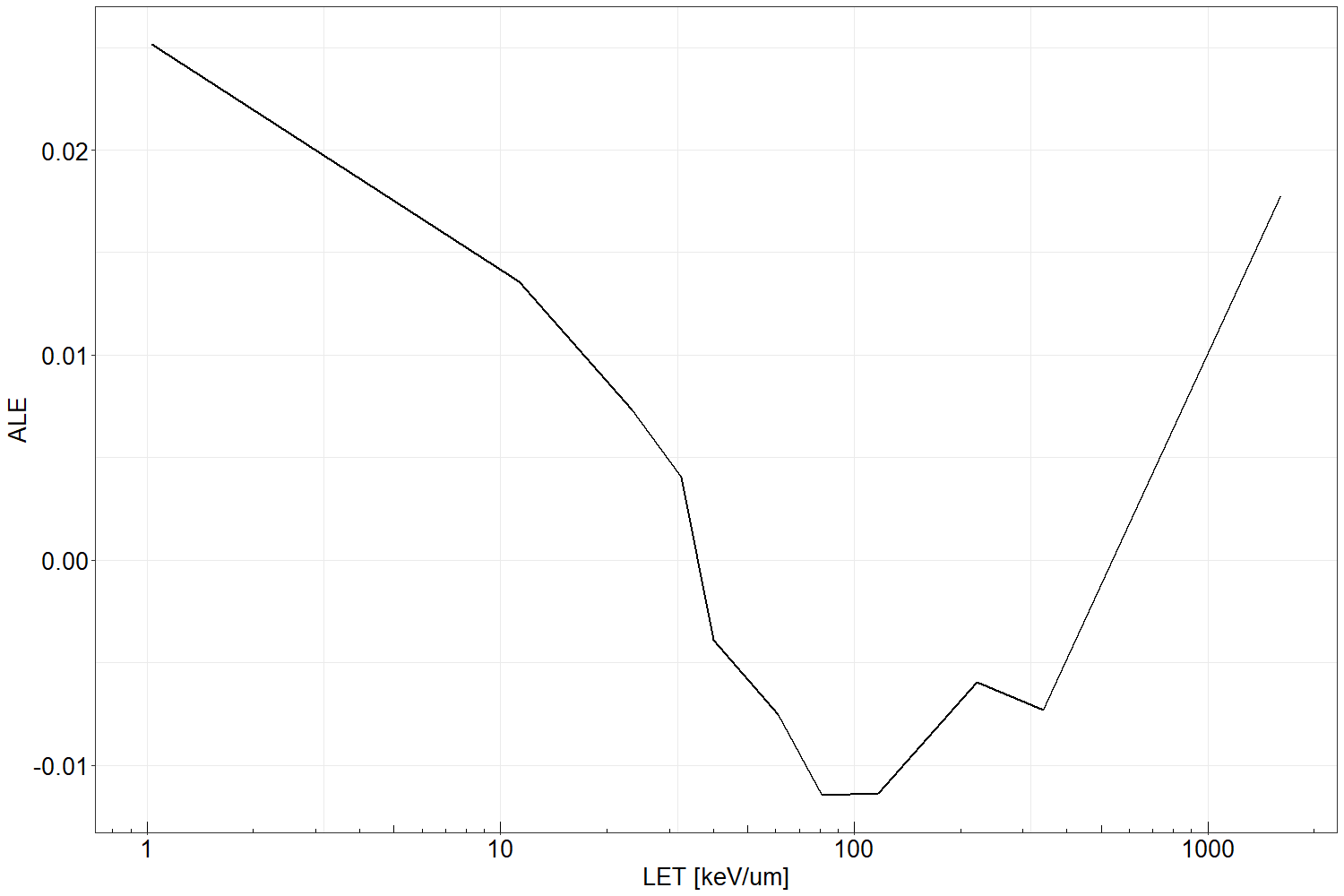}
    \caption{ALE plot of the effect of LET on the survival probability predicted by ANAKIN for an imparted dose of 2 Gy.}\label{FIG:ALES}
\end{subfigure}
\caption{ANAKIN global interpretability.}\label{FIG:VarImpAll}
\end{figure}

Figure\ref{FIG:VarImp} shows the global importance of all variables included in ANAKIN, calculated over the whole test dataset.
%To each input variable reported in the vertical axis, Figure \ref{FIG:VarImp} reports the global variable importance;
%The term global means that the importance is calculated over the whole test dataset.
The plot suggests that the dose is by far the most relevant parameter, followed by $\beta_\gamma$, \textit{LET} and \textit{Dose2} (the dose squared), which are all close together. This finding indicates that ANAKIN uses both physical and biological variables to predict the biological outcome. The \textit{Ions} \textit{Cells} variables, on which a categorical embedding has been performed, denotes the ion type and cell-line, respectively, and have both a high impact on ANAKIN.

\begin{comment}
\begin{figure}
    \centering
    \includegraphics[width=.9\textwidth]{Figure/VarImp.png}
    \caption{Importance \red{analysis} for the variables used by ANAKIN. The variable names reflect PIDE documentation \cite{friedrich2013systematic,friedrich2021update}. We also introduced the Dose2 as a new variable, representing the square of the dose.}\label{FIG:VarImp}
\end{figure}
\end{comment}

To a have a better understanding of ANAKIN global behavior, and in particular on the correlation between LET and the predicted survival, we calculated the ALE plot as described in Section \ref{SEC:MMInt} and reported in in Figure \ref{FIG:ALES} as a function of the LET. To obtain an unbiased effect, the ALE has \red{been} evaluated at the same dose of 2 Gy for all the experiments. The typical trend of the overkilling effect is clearly visible: Figure \ref{FIG:ALES} implies that small positive variations in the LET yields a clear negative variation in the cell survival, with a consequent increases of the RBE, up to 100 \kev, after which the cell survival starts \red{increasing} again as the LET increases, with therefore a decreases in the RBE.

%at increasing LET, the survival probability predicted by ANAKIN decreases, implying that the RBE increases, up to around 100 \kev, and then it starts growing again, with a consequent drop of the RBE.

%The ALE plot versus LET shows that ANAKIN can clearly predict the overkilling effect, even it is was not specifically trained for it. 

\begin{comment}
\begin{figure}
    \centering
    \includegraphics[width=.9\textwidth]{Figure/Ale_S.png}
    \caption{ALE plot of the effect of LET on the survival probability predicted by ANAKIN for an imparted dose of 2 Gy.}\label{FIG:ALES}
\end{figure}
\end{comment}

\begin{figure}
\begin{subfigure}[b]{\textwidth}
 \includegraphics[width=.9\textwidth]{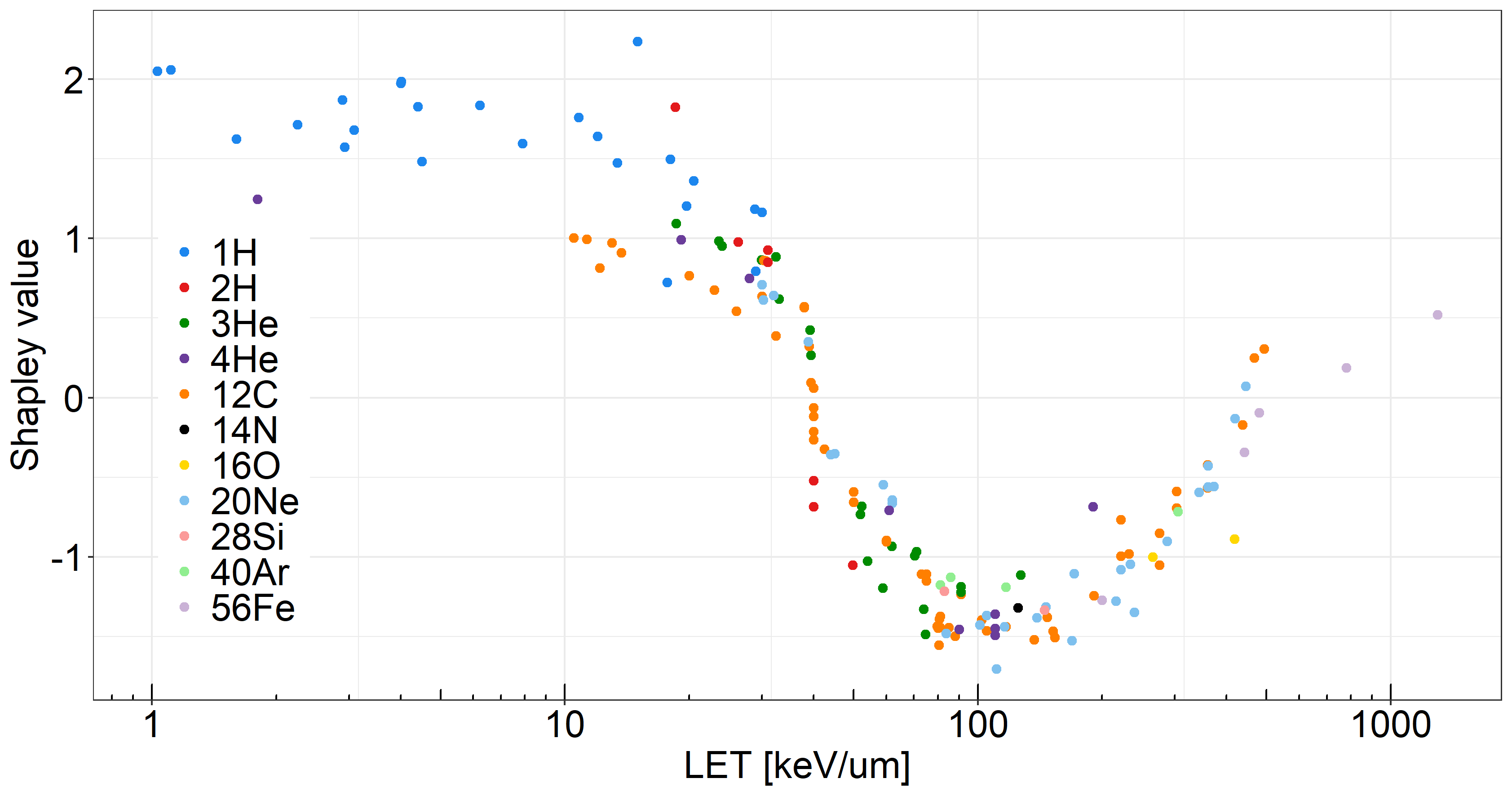}
    \caption{SHAP values for LET plotted against LET. All data are for a 2 Gy dose irradiation with different ions, ranging from protons to iron.}\label{FIG:ShapLET}
\end{subfigure}
\par\bigskip
\begin{subfigure}[b]{\textwidth}
\includegraphics[width=.9\textwidth]{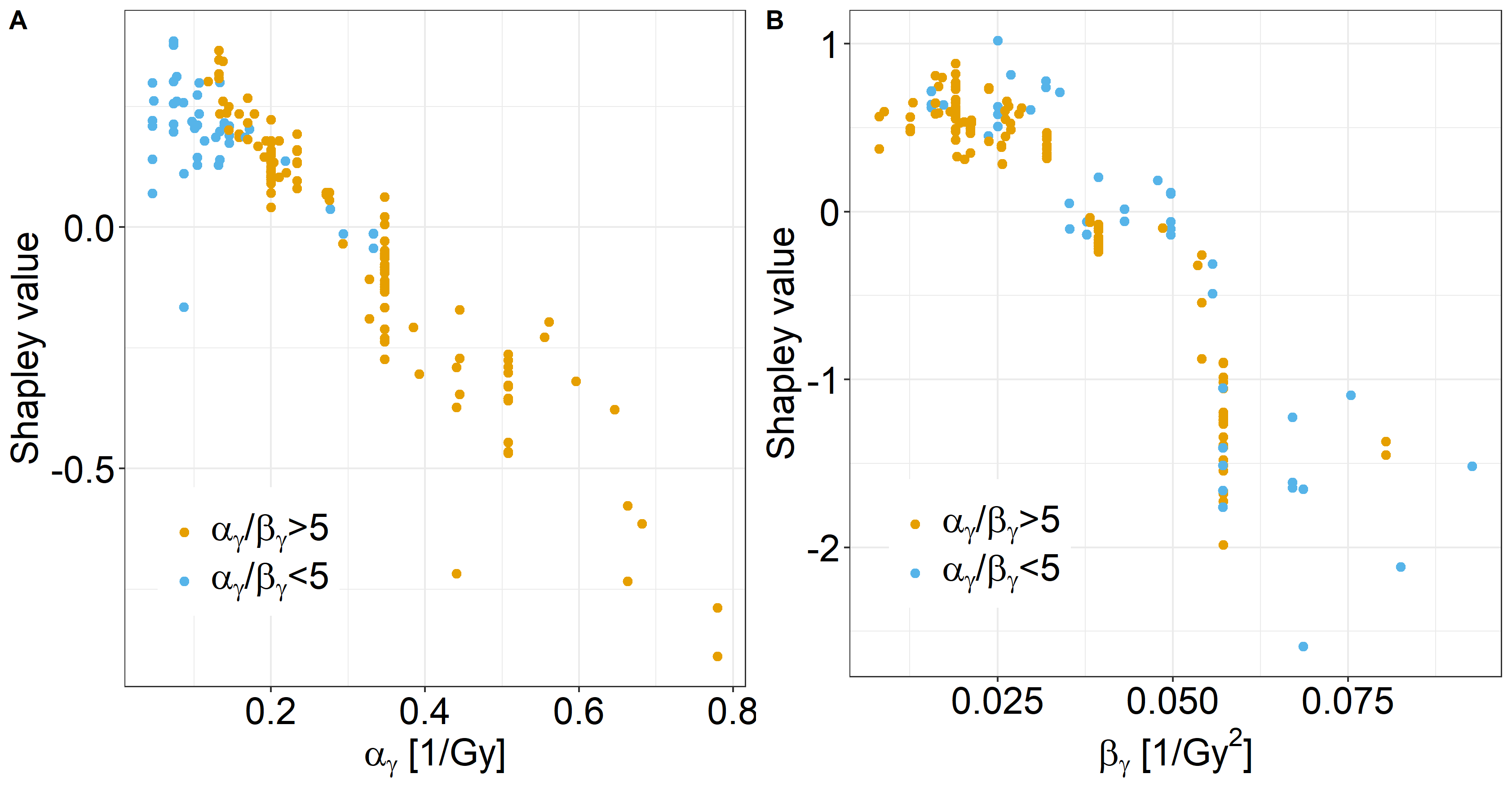}
    \caption{SHAP values calculated for $\alpha_\gamma$ plotted against $\alpha_\gamma$ (A) and $\beta_\gamma$ plotted against $\beta_\gamma$ (B). The data are divided into two groups, for \AB $>5$ (yellow dots) and \AB $< 5$ (blue dots). The data are for a dose of 2 Gy.}\label{FIG:ShapAB}
\end{subfigure}
\caption{ANAKIN SHAP value for LET and \AG and \BG parameters.}\label{FIG:ShapLETAll}
\end{figure}

\begin{figure}
\begin{subfigure}[b]{\textwidth}
 \includegraphics[width=.9\columnwidth]{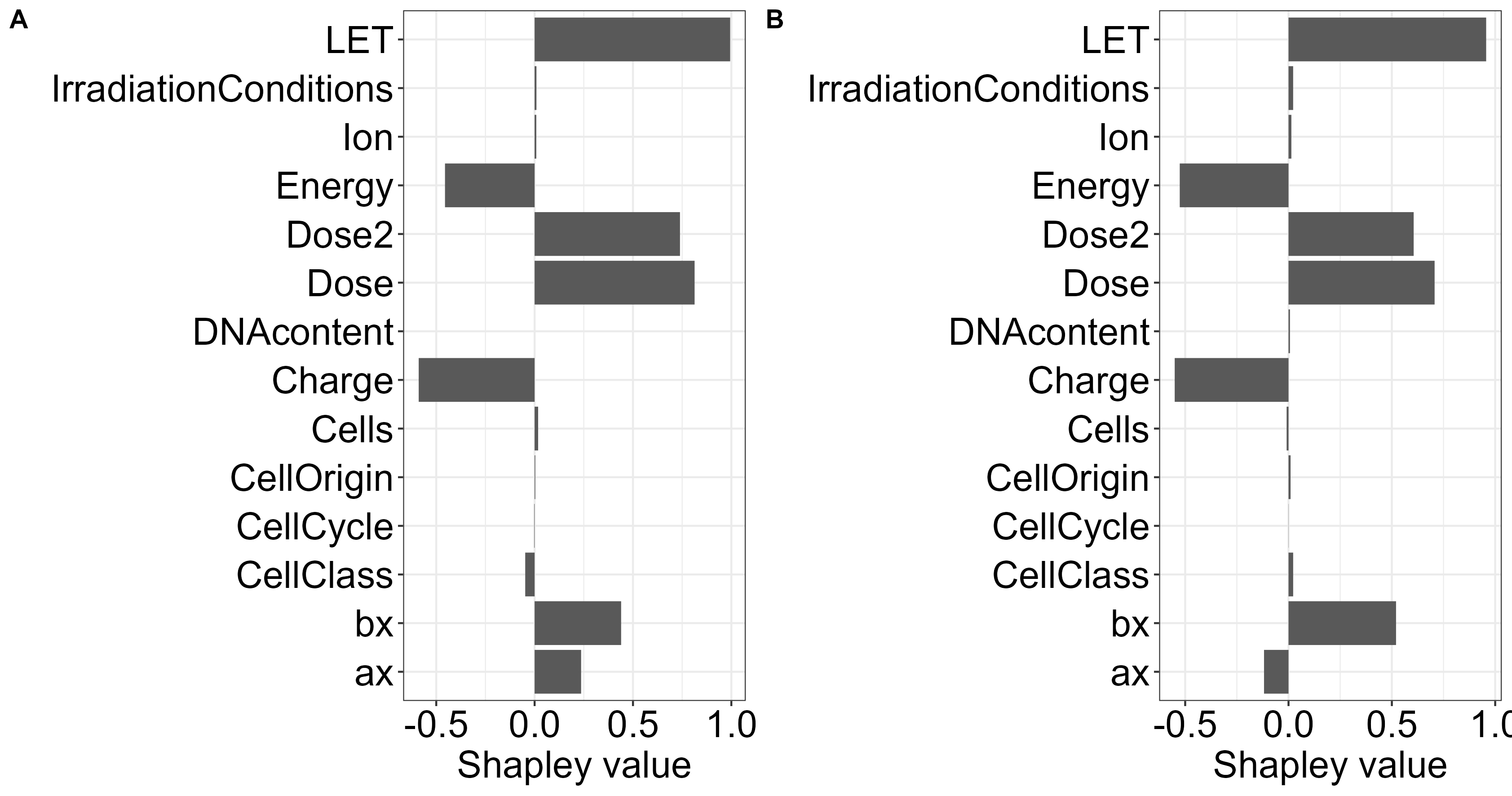}
         \caption{SHAP values calculated for the most relevant ANAKIN variables for protons of similar LETs ((A) 18 keV$/\mu$m and (B) 19 keV$/\mu$m). The dose has been set at 2 Gy.}
    \label{FIG:Shap2}
\end{subfigure}
\par\bigskip
\begin{subfigure}[b]{\textwidth}
 \includegraphics[width=.9\textwidth]{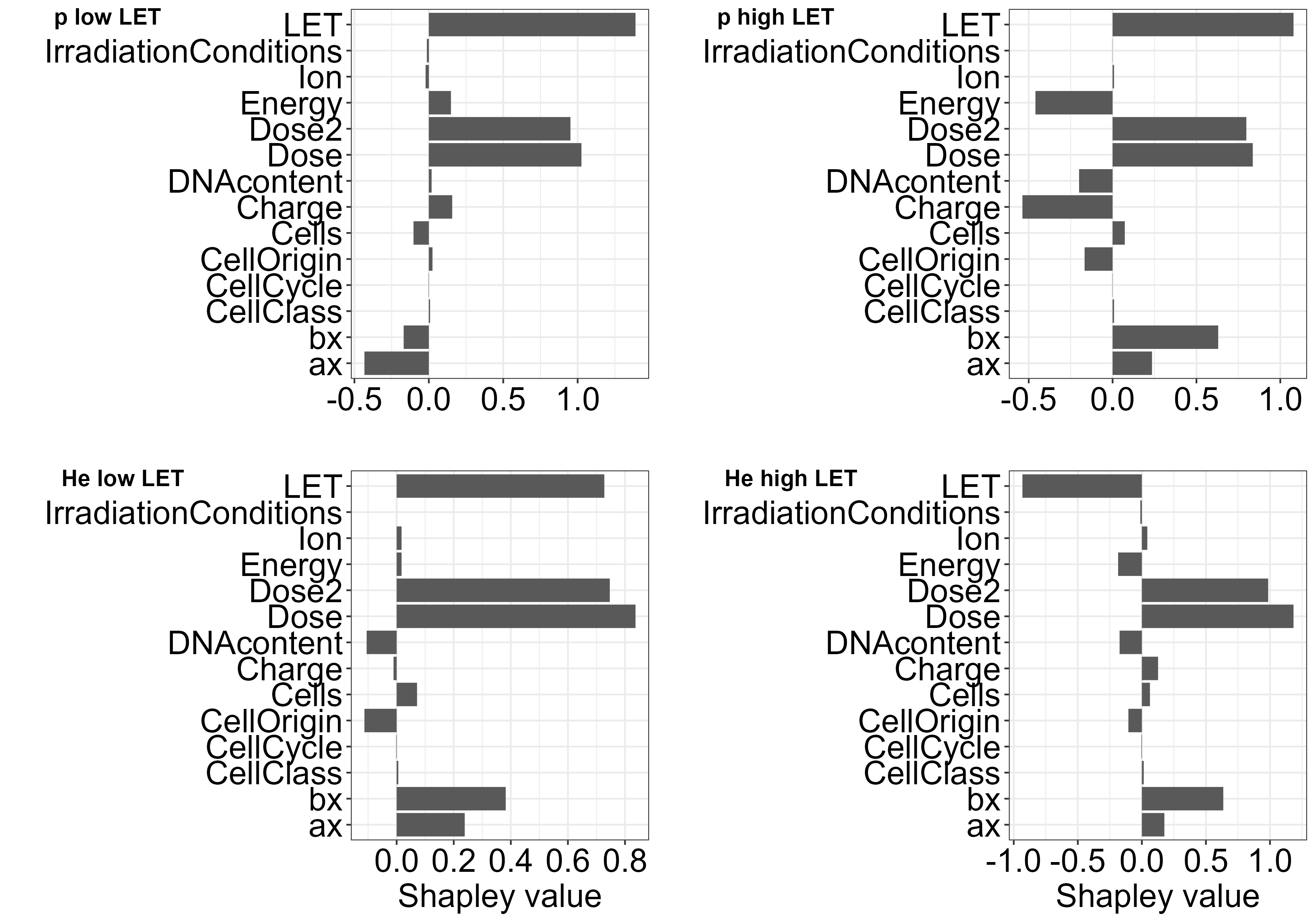}
            \caption{SHAP values calculated for the most relevant variables for the V79 cell line for protons at 4 keV$/\mu$m (A) and 28.8 keV$/\mu$m and helium at (C) 28 keV$/\mu$m and (D) 190 keV$/\mu$m. Dose has been set to 2 Gy.}\label{FIG:Shap4}
\end{subfigure}
\caption{ANAKIN SHAP values for different experiments.}\label{FIG:Shap2All}
\end{figure}

Figure \ref{FIG:ShapLET} reports the SHAP value for each experiment plotted against LET, considering a fixed dose of 2 Gy. Unlike the ALE plot, the SHAP value is a local technique, namely the SHAP is evaluated for each individual input variable,  and thus Figure \ref{FIG:ShapLET} shows the importance of LET in the overall cell survival assessment, evaluating it for each experiment. As for the ALE plot, the typical behavior of the overkilling effect emerges. Protons exhibit a high positive SHAP value, which remains almost constant up 15 \kev. As the LET rise above 15 \kev, protons shows a steep increase in the RBE, and the SHAP value for LET drops. In this region, especially for low-energy protons, the kinetic energy becomes more important than the LET to predict the cell-survival (Figure \ref{FIG:Shap4}). The SHAP value decreases steadily up to 100 \kev, after which it start increase again, reproducing the typical shape due to the overkill effect.

%ALE alpha beta ci sono già tante figure forse possiamo toglierla
%\begin{figure}
%    \centering
%    \includegraphics[width=.9\textwidth]{Figure/Ale_ab.png}
%    \caption{ALE plot of the effect of LET on the survival probability predicted by ANAKIN. ANAKIN prediction has been evaluated for an imparted dose of 2 Gy.}\label{FIG:ALES}
%\end{figure}

\begin{comment}
\begin{figure}
    \centering
    \includegraphics[width=.9\textwidth]{Figure/Shap_LET.png}
    \caption{SHAP values for LET plotted against LET. All data are for a 2 Gy dose irradiation with different ions, ranging from protons to iron.}\label{FIG:ShapLET}
\end{figure}
\end{comment}

Figure \ref{FIG:ShapAB} shows the SHAP value for \AG and \BG. The SHAP value for \AG shows that low \AG has a positive but almost equal importance to the model, but as \AG increases and \AB goes over 5 Gy the SHAP values linearly decreases to have at last negative high values.
%experiments are grouped by high and low \AB, respectively for \AB$ >5 $ Gy  and \AB$ < 5$ Gy.

For \AG below a certain threshold, that coincides for cell-lines with low \AB $<$ 5, the SHAP is positive, and then it starts decreasing linearly with \AG, reaching high negative values for high \AG and \AB. A similar trend is shown by the \BG SHAP values. For low \BG and high \AB, the SHAP value is positive, and then it begins to diminish. The data also indicate that the SHAP values for \BG show both higher variability and higher absolute values than those for \AG.

%On of the major differences between the SHAP value for \AG and \BG is that SHAP values for \BG show both higher variability and higher absolute values compared to the ones of \AG.
\begin{comment}
\begin{figure}
    \centering
    \includegraphics[width=.9\textwidth]{Figure/Shap_AB.png}
    \caption{SHAP values calculated for $\alpha_\gamma$ plotted against $\alpha_\gamma$ (A) and $\beta_\gamma$ plotted against $\beta_\gamma$ (B). The data are divided into two groups, for \AB $>5$ (yellow dots) and \AB $< 5$ (blue dots). The data are for a dose of 2 Gy.}\label{FIG:ShapAB}
\end{figure}
\end{comment}

%\begin{figure}
%    \centering
% \includegraphics[width=.9\columnwidth]{Figure/shap_p.png}
%         \caption{SHAP values calculated for the most relevant variables (variable names are as described in PIDE documentation, \cite{friedrich2013systematic,friedrich2021update}) for protons at comparable LETs ((A) 33 keV$/\mu$m and (B) 30 keV$/\mu$m). Dose has been set to the dose giving the RBE$_{10}$.}
%    \label{FIG:Shap_p}
%\end{figure}

\begin{comment}
\begin{figure}
    \centering
 \includegraphics[width=.9\columnwidth]{Figure/Shap2.png}
         \caption{SHAP values calculated for the most relevant ANAKIN variables for protons of similar LETs ((A) 18 keV$/\mu$m and (B) 19 keV$/\mu$m). The dose has been set at 2 Gy.}
    \label{FIG:Shap2}
\end{figure}
\end{comment}

Finally, we performed a comparison of experiments considering the SHAP values. Figure \ref{FIG:Shap2} shows the SHAP values for ANAKIN input features for two experiments performed with protons of similar LET of 18 \kev and 19 \kev for different cell lines. The corresponding \RBE{10} values of the two experiments are significantly different, being 1.2 and 2.7, respectively, as can be seen in Figure \ref{FIG:IonRBE} panel (A). ANAKIN outputs are extremely accurate for both experiments, being 1.13 and 2.5, with a MAE of 0.07 and 0.1 and a MAPE of 0.05 and 0.03, respectively. Figure \ref{FIG:Shap2} suggests that the only variables showing a significant difference between the two experiments are the \AG and \BG, as it should be since the two experiments have been performed over different cell-lines.

\begin{comment}
\begin{figure}
    \centering
    \includegraphics[width=.9\textwidth]{Figure/Shap4.png}
            \caption{SHAP values calculated for the most relevant variables for the V79 cell line for protons at 4 keV$/\mu$m (A) and 28.8 keV$/\mu$m and helium at (C) 28 keV$/\mu$m and (D) 190 keV$/\mu$m. Dose has been set to 2 Gy.}\label{FIG:Shap4}
\end{figure}
\end{comment}

Figure \ref{FIG:Shap4} compares the SHAP values for 4 different experiments, performed with either protons of helium of different LET (high or low). The rationale for the measurements selection is to test ANAKIN for   different ions and LET. %In particular, the following experiments have been compared: (A) protons at 12 \mev and 4 \kev, (B) protons at 0.9 \mev and 28.8 \kev, (C) helium at 5.9 \mev and 28 \kev and (D) helium at 0.45 \mev and 190 \kev.
Besides differences in the cell-lines specific parameters, focusing only on ion specific variables, it can be seen how LET and energies are treated significantly differently. The SHAP value for LET is high and positive for both protons datasets and for low-LET helium, while is negative for high-LET helium. The SHAP related to the beam energy is positive and close to 0 for both particles when the LET is low, it is negative and close to 0 for high-LET helium, and negative with a high absolute value for high-LET protons.

%\begin{figure}
%    \centering
% \includegraphics[width=.9\columnwidth]{Figure/shap_HL.png}
%        \caption{SHAP values calculated for the most relevant variables (variable names are as described in PIDE documentation, \cite{friedrich2013systematic,friedrich2021update}) for the V79 cell line for protons at (A) 4 keV$/\mu$m and (B) 28 keV$/\mu$m and helium at (C) 28 keV$/\mu$m and (D) 190 keV$/\mu$m. Dose has been set to the dose giving the RBE$_{10}$.}
%    \label{FIG:Shap_HL}
%\end{figure}

\section{Discussion}

The results reported in Section \ref{SEC:Res} show that ANAKIN produces accurate results over a wide range of biological endpoints, beams of different particle species and energies, with a consistent behavior for different error metrics. Despite the fact that \textit{logRMSE} is the most robust metric, being able to detect \red{discrepancies} between the predicted cell survival and the experiments at different doses, to be easily comparable to existing radiobiological model, most of the analysis of ANAKIN results has been conducted for RBE. As often in predictive analysis, the choice of the metric is of fundamental importance and strongly depends on the \red{specific} variable that the model is built to predict. In this particular case, the cell \red{survival curves} considered in the study have an extremely wide range, with corresponding RBE up to 6, and for this reason just a single metric cannot give a robust evaluation of the model accuracy. Experiments conducted with high LET radiation, are characterized by high RBE values, might have high absolute errors. However, the relative percentage error might be lower in such cases, being perhaps a more appropriate metric for experiments with high RBE. On the contrary, for experiments where the RBE is close to 1 , such as those conducted with high energy protons, the MAPE might be misleading, and the MAE might be a better tool. Since the choice of the most relevant metric is not always \red{trivial}, and depends on the chosen \red{endpoint}, our analysis was usually performed studying MAPE and MAE together. Nonetheless, for the sake of readability and to avoid giving an \red{excessive} amount of information, when \red{reasonable}, only the MAE metric is \red{reported} as it is considered to be, for the present case, more informative \red{as} compared to the MAPE.

MAPE and MAE \red{distributions} (Figure \ref{FIG:2x2RBEMAE} and able \ref{TAB:Error}) \red{show} that the errors for \RBE{10}, \RBE{1} and \RBE{50} are all sharply peaked around the average values with low deviation, denoting an overall consistent cell-survival prediction despite the extremely large \red{ range of LET and cell-lines} considered in the study. Further, it can be seen how ANAKIN is able to reproduce not only the average RBE, represented by the continuous splines, but also the high variability of the RBE across many LET and cell-lines. The validation of ANAKIK against \RBE{10} measurements (Figure \ref{FIG:2x2RBE}) shows the model accuracy. When the MAE is plotted against the LET, two key aspects emerge: (i) in the low-LET region, the MAE for \RBE{50} is slightly lower compared to high LET-region; (ii) the MAE for \RBE{10} and \RBE{1} remains almost constant in the whole LET range, with a slight drop at around 30 \kev. Notable enough, in the range 80-120 \kev, the experiments exhibit a \red{large} variation in RBE, nonetheless ANAKIN error does not seem to be affected by this huge variability with no evident increase in ANAKIN inaccuracy. This could go in the same direction as noted above, meaning that ANAKIN is able to predict RBE fluctuations at high LET.
%In this LET region, a precise \red{estimation of the LET of the experimental condition} could be not easy to perform so that the \red{larger} variability can be also a consequence of a \red{large} inaccuracy in the LET estimation. 
Further, a feature that support the potential of AI in modeling cell survival and RBE, is that ANAKIN predicts the overkill effect around 100 \kev, without any specific training.

ANAKIN \RBE{10} predictions show a slowly higher discrepancy from the experimental values in the low \BA region, again mostly for \RBE{50}, which corresponds to cell-lines with high \AB (Figure \ref{FIG:2x2RBE})
%it is clear that for low \BA ANAKIN underestimates the \RBE{10}. Same conclusion can be seen from the high MAE error in panel (D) in such region, with the error that significantly decraeses as \BA increases.
These cell-lines are extremely radiosensitive, and therefore are characterized by a \red{larger} experimental variability that reflected in the low accuracy of ANAKIN prediction. Further, less experiments have been performed for cell-line with high \AB, so that a higher error might simply be a natural fluctuation due to a lower statistics.

A specific analysis of single ions species prediction shows how ANAKIN accurately predicts cell survival over a wide range of ion species with very different LET, also guessing correctly the dependence of LET-RBE profiles on the ion type. For protons, ANAKIN is capable of reproducing the almost constant RBE at low-LET with a steep \red{increase} after 5 \kev, as well as the extremely high RBE at around 20 \kev. As shown\red{, for example} in \cite{missiaggia2022investigation}, currently used RBE models a often unable to accurately reproduce the RBE for \red{very} low energy protons.
%\red{Given the clear importance of protons to radiotherapeutic application in a clinical setting}, ANAKIN could then provide a robust and accurate tool to predict the biological dose over which perform \red{biological} treatment plan optimization, to improve the well-known constant value of 1.1 \red{currently} used in clinical application.
ANAKIN could then provide a robust and accurate tool to predict the RBE of clinical protons, thus allowing to develop TPS with a variable RBE instead of the fixed value of 1.1 currently used.

The comparison with experimental data acquired with helium and carbon ions show that ANAKIN prediction are accurate also for these species, even if exhibiting a \red{larger} variability on the errors. This is a direct consequence of the higher variability of RBE characterization \red{connected to these} two ions. These findings suggest that ANAKIN could provide an invaluable tool for predicting RBE for heavy ions, where \red{it is commonly accepted that a } constant value cannot be used in their clinical applications. 
%As also mentioned above, a clear overkill effect is predicted by ANAKIN in the high LET region. 
%The accuracy of ANAKIN is also reflected into the error distribution. Protons show few experiments characterized by errors, due mainly to the low absolute value of RBE characterizing protons, which therefore are more likely to induce high MAE errors.

%Error distribution for monoenergetic and SOBP is surprising as first, as it could be expected that mono-energetic ion beams would be characterized by lower errors.
The error distribution for monoenergetic beams is lower than for SOBP, as indicated by the main peak of the MAE distributions (Figure \ref{FIG:MAPESOBP}), but it is broader. We hypothesize that this behavior could be due to the fact that for a monoenergetic beam, an inaccurate LET estimation can result in a significantly different RBE prediction. Overall, ANAKIN is able to accurately predict RBE value for both monoenergetic and SOBP, without the need of adding ad hoc \red{adaptations}.

Also \RBE{\alpha} and \RBE{\beta} show a good accuracy between predicted and experimental values. Both $\alpha$ and $\beta$ errors seems to remain constants over the whole range of LET, whereas an analysis of the $\alpha$ variability as a function of \BA shows that for high \AB cell-lines the estimation of $\alpha$ is subject to higher uncertainty. There is a clear underestimation of $\alpha$ for high \AB cell-line, which directly translated into the high RBE error in these cell-lines, as already discussed above. Similar conclusions can be drawn for $\beta$ with a slightly higher error in the high \BA region. These results point out that ANAKIN is able to reproduce not only $\alpha$, but also $\beta$, which is typically subject to an extremely high uncertainty, as shown for instance by low accuracy of many models in reproducing $\beta$ variability, \cite{pfuhl2022comprehensive}. In addition, $\beta$ is predicted to be dependent on the radiation quality, as shown by the trend of the experimental data and \red{in contrast to} many other existing radiobiological models.

\subsection{Comparison with MKM and LEM}

To further test ANAKIN capability and appreciate its accuracy, we compare its results with predictions from the two radiobiological models currently used in the clinics (MKM and LEM). 

The finding of this comparison indicate that ANAKIN has an overall higher accuracy than both the MKM and the LEM in all the metrics. The MKM performs slightly better than the LEM, but this result could be related to the fact that we had to use LEM III, instead of the latest version LEM IV, which was not available. This hypothesis is supported by the results reported in \cite{pfuhl2022comprehensive}, which shows that LEM IV has significantly better accuracy than LEM III. 

ANAKIN error distribution is significantly less broad that both the MKM and LEM distributions, and its maximum error is lower. ANAKIN has not been specifically trained to predict the two cell-lines selected for the comparison (i.e. V79 and HSG), but on a wide range of different cell-lines available onn PIDE. %The MKM shows accurate results mostly for the HSG cell-line, probably because it has been vastly tested in literature over a high numbers of cell survival experiments.

The analysis performed on several experiments suggests that ANAKIN is more accurate than both the MKM and LEM. Overall, ANAKIN shows a lower error than the MKM and LEM, and even when its prediction is less accurate than the other two models, the maximum error is lower than those obtained with the other two.

%as well as the  he highest discrepancy in the errors is significantly higher when ANAKIN has a lower error than the case where on the opposite either the MKM or the LEM has a lower error.

\subsection{Explainable Artificial Intelligence}

%Despite the fame of being black box, some advanced and extremely powerful ML and DL models are instead easily interpretable. Particularly, tree-based ML models are extremely flexible and a good comprehension of the relations learn between the input variable and prediction.

The global variable importance study presented in Figure \ref{FIG:VarImp} shows how both biological and physical variables are efficiently used by ANAKIN to predict cell survival. The dose is the most important variable, as expected. The analysis also identifies the square of the dose as a relevant variable, which is also reasonable as the quadratic relation between the survival and the dose is widely used in many RBE models. Concerning the physical variables, LET is considered to be more important than the ion kinetic energy. 
%further insights into this will emerge in subsequent analysis. 

For the biological variables, \AG and \BG are among the most relevant input parameters together with the cell-line. Our hypothesis is that ANAKIN uses these three \red{variables} to understand how a specific cell-line responds to ionizing radiation. 

The variables over which a Deep Embedding was performed, namely \textit{Ion}, \textit{Cells} and \textit{CellCycle}, are also relevant to the model predictions, suggesting that such advanced DL based embedding has been able to uncover important information.

The same effect emerges analyzing the dependence of the SHAP value from the LET. For protons, ANAKIN gives approximately the same positive and high importance to LET up to 15 \kev. In this region, protons shows an almost constant RBE, and ANAKIN recognizes this behavior by giving the same importance to different values of LET.

%The SHAP value for \AG shows that low \AG has a positive but almost equal importance to the model, but as \AG increases and \AB goes over 5 Gy the SHAP values linearly decreases to have at last negative high values. This trend is similar to that already observed ???, where a high RBE variability can be detected for cell lines with high \AB. 

The association between high \AG values and high negative importance in Figure \ref{FIG:ShapAB} reflects the fact that, for highly radiosensitive cell--lines, the \AG value, that describe the contribution of single track, should be more important than in cell-lines with lower \AB.

The SHAP value could be also extremely important in understanding \red{which} variables lead to a certain RBE. To support this hypothesis, we compared two experiments conducted with protons of comparable energies, namely 18 \kev and 19 \kev. The results indicate that ANAKIN can correctly predict a significant variability in the RBE, that in this case stemmed from the fact that different cell lines were considered, as pointed out by the SHAP values for the \AG and \BG parameters.

To investigate how different physical variables affect ANAKIN outcomes, we considered proton and helium beams of different energies. We found that LET has always a positive high impact only for high-LET helium, as this beam was the only one with an LET in the overkill region. Therefore it might be concluded that the LET is used by ANAKIN in a significantly different manner when an overkill effect is expected. The SHAP value for the beam kinetic energy is positive for protons and helium at low LET, whereas \red{it} is negative for the high LET beams, which have have extremely low energies (below 1 \mev), corresponding to depth downstream of the Bragg peak. The SHAP analysis indicates that the kinetic energy is much more important for protons than helium at low values. This behavior can be due to the fact that low-energy helium ions have an LET above the overkill threshold, and thus the main information to predict cell survival is carried by LET. Overall, it seems that ANAKIN is able to use jointly LET and kinetic energy to accurately predicts the cell survival fraction.

Advanced XAI techniques have been applied to understand what variables are relevant to ANAKIN predictions, as well as to show how specific biological features observed in experimental data, such as the overkilling effect \red{at high LET} and the variable $\beta$ coefficients, are reproduced by ANAKIN. The implementation of such behaviors is non-trivial in a purely mathematical model and it represent thus a strength of ANAKIN. Furthermore, these XAI techniques can play a major role in clinical applications since they allow the interpretation of ANAKIN prediction but also the understanding on how reliable the given prediction could be. It is worth stress that, \red{one of the major limitation in biophysical modelling of radiation effects, both for curative and radioprotective purposes, is exactly on the uncertainties estimation}. Although an AI-based approach is not derived from mechanistic considerations, unlike existing radiobiological models, and thus cannot provide a validation of these mechanism, on the other hand its power in processing and filtering the data dependencies can help to reveal features hidden in the data, that on \red{their} turn can drive further comprehension of the phenomenon.

\section{Conclusions}
The present paper presents the AI-based model ANAKIN, for predicting the survival probability of various cell lines exposed to \red{different types of} radiation. The findings contained in this paper, proves that a single model is able to predict the behavior of different ion species, without the need of specifically train the model on data relative to a single ion.  %ANAKIN is extremely general, being able to include several different ions as well as many different cell lines.
Although the main motivation for developing ANAKIN is to apply it in particle therapy, the model accuracy in predicting the biological effect of extremely high LET events could extend its application in other fields, such as space radioprotection.
%hint to \red{its possible efficient application also in other} fields, such as space radioprotection.

The analysis described here indicates that ANAKIN is able to accurately predict cell survival and RBE over a wide range of different cell-lines and ions type. Higher uncertainties and errors emerges for cell--lines characterized by low \AB and LET in the range from 20 to 150 \kev. These uncertainties reflects the uncertainties in the experimental data, on which ANAKIN has been trained on. In fact, cell-lines with high \AB as well as experiments with high LET beams show a higher variability of RBE. 
%Therefore, considered that a single model has been trained and that ANAKIN is capable to reproduce a series of extremely different experiments, above uncertainties seem natural. However, a deeper investigation of these aspects must be certainly conducted in future research.

When compared with two of the mostly used radiobiological models, namely the MKM and LEM III, ANAKIN showed in average more accurate predictions. The gap between the models could be smaller if the latest versions of the MKM and LEM become available in literature.
%The versions used for the MKM and LEM were not the  advanced versions of the MKM and LEM might give better results.

Although purely data-driven models are often considered to be less powerful than mechanistic models, ML and DL have the advantage of being extremely flexible. This is supported by the fact that ANAKIN predicts both the overkill effect and the variable $\beta$ into the MKM. On the contrary, in mechanistic model \textit{ad hoc} correction terms must typically be added to include above effects.

%The present analysis shows how, on the contrary, both effects have already been precisely modeled by ANAKIN. This aspect could in turn be a major strength of ANAKIN.

%Among the most relevant features of ANAKIN, beside its accuracy, there is its flexibility.
The modular structure of ANAKIN makes very easy to include advanced features. The most relevant example is the implementation of a radiation quality description different from the classical LET, such as \red{microdosimetric or nanodosimetric quantities}, as well as the coupling of ANAKIN with a mechanistic RBE model.

In conclusion, we showed that ANAKIN is an intuitive and understandable model, that demonstrates high accuracy in predicting cell survival and RBE. Any prediction given by ANAKIN can be analyzed into details, so the contribution of each input variable can be precisely assessed. Several advanced techniques of XAI can be used either to understand if a well-known biological or physical phenomena, such as the overkill effect or LET dependent $\beta$, has been included in ANAKIN, but also to gain further insight and unveil existing correlations between \red{different} variables.
\red{While AI has been broadly employed in radiotherapy treatment planning, either for physical dose optimization or image segmentation, ANAKIN is the first application on radiobiological calculations, which may open the possibility to use for the first time an AI-based model to \emph{biological} treatment planning, i.e. the optimization of the dose delivery with explicit consideration of a voxel-dependent RBE. This problem is typically extremely heavy computationally and strongly linked to uncertainties, two features where a model like ANAKIN may be of the outermost advantage.}

\cleardoublepage
%\nocite{*}
\bibliographystyle{apalike}
\bibliography{bib}

\end{document}